\documentclass[twocolumn,10pt,aps,pre,longbibliography,superscriptaddress]{revtex4-2}

\usepackage{graphicx}
\usepackage{amsmath}
\usepackage{amsfonts}
\usepackage{amssymb}
\usepackage{color}
\usepackage{float}
\usepackage{mathrsfs}
\usepackage{multirow}

\begin{document}

\title{Multisoliton interactions approximating the dynamics of breather solutions}

\author{D.\,S.~Agafontsev}
\affiliation{Shirshov Institute of Oceanology of RAS, 117997, Moscow, Russia}
\affiliation{Department of Mathematics, Physics and Electrical Engineering, Northumbria University, Newcastle upon Tyne, NE1 8ST, United Kingdom}

\author{A.\,A.~Gelash}
\affiliation{Laboratoire Interdisciplinaire Carnot de Bourgogne (ICB), UMR 6303 CNRS -- Université Bourgogne Franche-Comté, 21078 Dijon, France}

\author{S.~Randoux}
\affiliation{Univ. Lille, CNRS, UMR 8523 - PhLAM -  Physique des Lasers Atomes et Mol\'ecules, F-59000 Lille, France}

\author{P.~Suret}
\affiliation{Univ. Lille, CNRS, UMR 8523 - PhLAM -  Physique des Lasers Atomes et Mol\'ecules, F-59000 Lille, France}

\begin{abstract}
Breather solutions are considered to be generally accepted models of rogue waves. 
However, breathers are not localized, while wavefields in nature can generally be considered as localized due to the limited spatial dimensions. 
Hence, the theory of rogue waves needs to be supplemented with localized solutions which evolve locally as breathers. 
In this paper, we present a universal method for constructing such solutions from exact multi-soliton solutions, which consists in replacing the plane wave in the dressing construction of the breathers with a specific exact $N$-soliton solution converging asymptotically to the plane wave at large number of solitons $N$. 
On the example of the Peregrine, Akhmediev, Kuznetsov-Ma and Tajiri-Watanabe breathers, we show that the constructed with our method multi-soliton solutions, being localized in space with characteristic width proportional to $N$, are practically indistinguishable from the breathers in a wide region of space and time at large $N$. 
Our method makes it possible to build solitonic models with the same dynamical properties for the higher-order rational and super-regular breathers, and can be applied to general multi-breather solutions, breathers on a nontrivial background (e.g., cnoidal waves) and other integrable systems. 
The constructed multi-soliton solutions can also be generalized to capture the spontaneous emergence of rogue waves through the spontaneous synchronization of soliton norming constants, though finding these synchronizations conditions represents a challenging problem for future studies. 
\end{abstract}

\maketitle


\section{Introduction}
\label{Sec:Intro}

The phenomenon of rogue waves -- unusually large waves that appear suddenly from moderate wave background -- has been extensively studied in recent years. 
A number of mechanisms have been suggested to explain their emergence, see e.g. reviews~\cite{kharif2003physical,dysthe2008oceanic,onorato2013rogue,dudley2019rogue}, with one of the most general ideas being that rogue waves could be related to breather-type solutions of the underlying nonlinear evolution equations~\cite{dysthe1999note,osborne2000nonlinear,osborne2010nonlinear,shrira2010makes}. 
A particular example of a nonlinear mathematical model suitable for the description of rogue waves is the one-dimensional nonlinear Schr{\"o}dinger equation (1D-NLSE) of the focusing type,
\begin{equation}\label{NLSE}
	i\psi_t + \frac{1}{2}\psi_{xx} + |\psi|^2 \psi = 0,
\end{equation}
which describes evolution of a narrowband signal in a weakly nonlinear media and is widely applicable in different fields of studies ranging from nonlinear optics to hydrodynamics and Bose-Einstein condensates~\cite{kivshar2003optical,pelinovsky2008book,OsborneBook2010}. 
Several exact solutions of this equation have been suggested as candidates of rogue waves, including the Peregrine~\cite{peregrine1983water}, Akhmediev~\cite{akhmediev1986modulation} and Kuznetsov-Ma~\cite{kuznetsov1977solitons,kawata1978inverse,ma1979perturbed} breathers, which mathematically represent special cases of the Tajiri-Watanabe breather~\cite{its1988exact,tajiri1998breather}. 
Taking specific and carefully designed initial conditions, these solutions were reproduced in well-controlled experiments performed in different physical systems~\cite{kibler2010peregrine,chabchoub2011rogue,clauss2011formation,bailung2011,kibler2012observation,frisquet2013collision,chabchoub2019drifting}.

Meanwhile, in many physically relevant situations, the wavefields are localized due to the limited spatial dimensions. 
In such systems, breather solutions cannot formally exist, since they have finite background boundary conditions and are not localized. 
This means that the observed breather-like dynamics must be achieved through a combination of localized structures such as nonlinear dispersive waves and solitons, and the theory of rogue waves needs to be supplemented with localized solutions which evolve locally as breathers. 
Additionally, the similarity between certain breather profiles and interacting solitons~\cite{gelash2018strongly,agafontsev2021rogue,agafontsev2021extreme,chabchoub2021peregrine} observed in integrable systems, as well as the fission of breather solutions into solitons~\cite{chowdury2023rogue,chowdury2022higher} in nonintegrable systems, also imply a close relationship between breathers and solitons.

In our paper, we aim to address these observations by presenting a universal method for modeling breather-like dynamics with exact multi-soliton solutions. 
Our approach is based on the integrability of the 1D-NLSE in terms of the \textit{inverse scattering transform} (IST) theory, and, more specifically, on the dressing method also known as the Darboux transformation. 
Namely, breathers are constructed by ``dressing'' a plane wave solution $\psi = A\,e^{i|A|^{2}t}$ of Eq.~(\ref{NLSE}) with solitons. 
We modify this construction by replacing the plane wave with a specific exact $N$-soliton solution~\cite{gelash2021solitonic}, which converges asymptotically to the plane wave at large number of solitons $N$. 
As a result, we obtain exact multi-soliton solutions which approximate breathers locally very well even for a rather small number of solitons $N\sim 10$. 
At large $N$, the constructed solutions are practically indistinguishable from breathers in a wide region of space and time. 
We believe that our method can be applied straightforwardly to general multi-breather solutions, breathers on a nontrivial background (e.g., cnoidal waves), and other integrable systems including vector breathers~\cite{kedziora2014rogue,baronio2014vector,chen2019rogue,che2022nondegenerate,gelash2023vector,hoefer2023kdv}. 

The paper is organized as follows. 
In Section~\ref{Sec:Methods}, we discuss the dressing method procedure, the multi-soliton and multi-breather solutions, and the solitonic model of the plane wave~\cite{gelash2021solitonic}. 
In Section~\ref{Sec:Results}, we construct solitonic models for the Peregrine, Akhmediev, Kuznetsov-Ma and Tajiri-Watanabe breathers, and, on the example of the Peregrine breather, numerically verify that these models approximate breathers exceptionally well in a wide region of space and time.  
The final Section~\ref{Sec:Conclusions} contains conclusions and discussions. 
In Appendixes~\ref{Sec:App:A} and~\ref{Sec:App:B}, we demonstrate in detail our solitonic models for the Akhmediev, Kuznetsov-Ma and Tajiri-Watanabe breathers, and also generalize our method for the higher-order rational~\cite{akhmediev2009rogue} and super-regular~\cite{zakharov2013nonlinear,gelash2014superregular,kibler2015superregular} breathers. 
In Appendix~\ref{Sec:App:C}, we argue that the universality of our method comes from the locality and continuity of the dressing procedure. 


\section{Dressing method construction}
\label{Sec:Methods}


\subsection{Integrability and the dressing method}
\label{Sec:Methods:A}

The 1D-NLSE belongs to a class of nonlinear partial differential equations (PDEs) integrable by means of the IST method. 
The latter is based on the representation of a PDE as a compatibility condition of an over-determined auxiliary linear system -- the Lax pair~\cite{ablowitz1981solitons,novikov1984theory}. 
For the case of the 1D-NLSE, the Lax pair is known as the Zakharov-Shabat (ZS) system~\cite{zakharov1972exact} for a two-component vector wave function $\mathbf{\Phi}(x,t,\lambda) = (\phi_1,\phi_2)^{T}$,
\begin{eqnarray}
	\mathbf{\Phi}_{x} &=& \begin{pmatrix} -i \lambda & \psi \\ -\psi^* & i \lambda \end{pmatrix}\mathbf{\Phi},
	\label{ZSsystem1}\\
	\mathbf{\Phi}_t &=& \begin{pmatrix}\ -i\lambda^2 + \frac{i}{2} |\psi|^2 & \lambda \psi + \frac{i}{2} \psi_x \\ -\lambda \psi^* + \frac{i}{2} \psi^*_x & i\lambda^2 - \frac{i}{2} |\psi |^2 \end{pmatrix} \mathbf{\Phi},
	\label{ZSsystem2}
\end{eqnarray}
where the star stands for the complex conjugate and Eq.~(\ref{NLSE}) is obtained as a compatibility condition $\mathbf{\Phi}_{xt} = \mathbf{\Phi}_{tx}$. 
The first equation of the ZS system can be rewritten as an eigenvalue problem for a complex-valued spectral parameter $\lambda = \xi + i \eta$, 
\begin{equation}\label{ZSsystem1-eigenvalue}
	\widehat{\mathcal{L}}\mathbf{\Phi} = \lambda \mathbf{\Phi},
	\quad
	\widehat{\mathcal{L}} = i \begin{pmatrix}\ 1 & 0 \\ 0 & -1 \end{pmatrix}\frac{\partial}{\partial x} - i\begin{pmatrix}\ 0 & \psi \\ \psi^* & 0 \end{pmatrix}.
\end{equation}
Similarly to the Schr{\"o}dinger operator in quantum mechanics~\cite{landau1958quantum}, the scattering problem~(\ref{ZSsystem1-eigenvalue}) for the ZS operator $\widehat{\mathcal{L}}$ and wave function $\mathbf{\Phi}$ can be introduced, in which the wavefield $\psi$ of the 1D-NLSE plays the role of a potential. 
For localized potentials, this problem has bounded solutions $\mathbf{\Phi}$ for real-valued spectral parameter $\lambda = \xi\in\mathbb{R}$ (continuous spectrum), and also for a finite number of discrete points $\lambda_{n} = \xi_{n} + i\eta_{n}$, $\eta_{n}>0$, $n=1,...,N$ (discrete spectrum). 
Note that, without loss of generality, we consider only the upper half of the $\lambda$-plane, $\eta = \mathrm{Im}\,\lambda \ge 0$. 

Most importantly, the potential $\psi(x,t)$ turns out to be in one-to-one correspondence with the so-called \textit{scattering data} -- a combination $\{\lambda_{n}, \rho_{n}(t), r(\xi,t)\}$ of the discrete spectrum points $\lambda_{n}$, associated with them coefficients $\rho_{n}(t)$, and reflection coefficient $r(\xi,t)$ representing the continuous spectrum -- and this scattering data changes trivially over time. 
In particular, the 1D-NLSE evolution preserves the discrete spectrum, $\partial_{t}\lambda_{n}=0$, while $\rho_{n}(t)$ and $r(\xi,t)$ evolve exponentially like the Fourier harmonics in the linear waves theory. 
These properties make it possible to fundamentally solve the Cauchy initial value problem for the 1D-NLSE by finding the scattering data from the solution of the scattering problem~(\ref{ZSsystem1-eigenvalue}), calculating its evolution in time, and recovering the potential $\psi$ with the inverse scattering transform. 
Note, however, that both the direct transformation to the scattering data $\{\lambda_{n}, \rho_{n}(t), r(\xi,t)\}$ and the inverse transformation to the potential $\psi(x,t)$ represent highly nontrivial nonlinear problems. 
In particular, the IST is done by solving the integral Gelfand-Levitan-Marchenko equations~\cite{novikov1984theory} and can be calculated analytically only in special cases, asymptotically at large time $t\to\pm\infty$, or in the semi-classical approximation~\cite{lewis1985semiclassical,jenkins2014semiclassical}. 

Like the Fourier spectrum in the linear waves theory, the scattering data is used to characterize the potential $\psi(x,t)$: the reflection coefficient $r(\xi,t)$, representing the continuous spectrum $\lambda = \xi\in\mathbb{R}$, describes the nonlinear dispersive waves, while the discrete eigenvalues $\lambda_{n}$ together with the coefficients $\rho_{n}(t)$ correspond to solitons. 
In particular, the eigenvalues $\lambda_{n}=\xi_{n}+i\eta_{n}$ contain information about the (invariant in time) soliton amplitudes $a_{n}=2\eta_{n}$ and velocities $v=-2\xi_{n}$, while the coefficients $\rho_{n}(t)$ -- about their (evolving) position in space and complex phase. 
In the present paper, we pay special attention to reflectionless potentials $r(\xi,t)=0$, i.e., solutions of the 1D-NLSE that contain only solitons. 

The dressing method (DM)~\cite{novikov1984theory,matveev1991darboux}, also known as the Darboux transformation~\cite{akhmediev1991extremely,matveev1991darboux}, is a simplified version of the conventional IST approach. 
It represents an algebraic scheme for constructing new exact solutions to integrable PDEs by ``dressing'' a bare solution, and is often used to find general multi-soliton and multi-breather solutions of the 1D-NLSE, see e.g.~\cite{zakharov1978relativistically,akhmediev1991extremely,akhmediev2009extreme,gelash2014superregular}. 
Here we discuss the DM optimized for dressing with solitons. 
Note that for this task the DM requires far fewer arithmetic operations than the standard IST method, that makes it favorable when the number of solitons is large~\cite{gelash2018strongly,gelash2019bound,tarasova2023properties}.

The DM is performed at a specific moment of time which we fix to zero $t=0$ for simplicity. 
It starts from an initial background potential $\psi_{0}(x,t)$ (bare solution) and the corresponding $2\times 2$ matrix solution $\mathbf{\Phi}_{0}(x,t,\lambda)$ of the ZS system (wave function) constructed from two linearly independent solutions of Eqs.~(\ref{ZSsystem1})-(\ref{ZSsystem2}). 
At the $n$-th step of the recursive procedure, the potential $\psi_{n}(x,t)$ dressed with $n$ solitons is constructed from $\psi_{n-1}(x,t)$ and the corresponding wave function $\mathbf{\Phi}_{n-1}(x,t,\lambda)$ as
\begin{eqnarray}
	\psi_{n}(x,t) = \psi_{n-1}(x,t) + 2i(\lambda_n-\lambda^*_n)\frac{q^*_{n1}q_{n2}}{|\mathbf{q_n}|^2}.
	\label{psi_n}
\end{eqnarray}
In this relation, vector $\mathbf{q}_{n}=(q_{n1},q_{n2})^{T}$ is determined via $\mathbf{\Phi}_{n-1}$, the soliton eigenvalue $\lambda_{n}$, and an arbitrary constant $C_{n}\in\mathbb{C}\backslash \{0\}$ which we call the norming constant,
\begin{eqnarray}
	\mathbf{q}_{n}(x,t) = [\mathbf{\Phi}_{n-1}(x,t,\lambda^*_n)]^{*} \cdot \left(\begin{array}{c} C_n^{-1/2} \\C_n^{1/2} \end{array}\right).
	\label{qn}
\end{eqnarray}
The corresponding wave function $\mathbf{\Phi}_{n}(x,t,\lambda)$ is calculated using the dressing matrix $\boldsymbol{\sigma}^{(n)}(x,t,\lambda)$, 
\begin{eqnarray}
	\mathbf{\Phi}_{n}(x,t,\lambda) &=& \boldsymbol{\sigma}^{(n)}(x,t,\lambda)\cdot \mathbf{\Phi}_{n-1}(x,t,\lambda), \label{dressing-Psi}\\
	\sigma^{(n)}_{ml}(x,t,\lambda) &=& \delta_{ml} + \frac{\lambda_n-\lambda_n^*}{\lambda - \lambda_n}\frac{q_{nm}^{*}q_{nl}}{|\mathbf{q_n}|^2},
	\label{dressing-matrix}
\end{eqnarray}
where $m,l=1,2$ and $\delta_{ml}$ is the Kronecker delta.
The outcome of the dressing by $N$ solitons can be written via the ratio of two determinants, 
\begin{eqnarray}
	&&\psi_{N}(x,t) = \psi_{0} - 2i\frac{\mathrm{det} \mathbf{P}}{\mathrm{det} \mathbf{Q}}, 
	\quad Q_{kj}=\frac{(\mathbf{\tilde{q}}_{k}\cdot \mathbf{\tilde{q}}^*_{j})}{\lambda_{k} - \lambda^*_j}, \nonumber\\
	&&\mathbf{P}=
	\left(\begin{array}{cc}
	        0 & \begin{array}{ccc}
	              \tilde{q}_{12} & \cdots & \tilde{q}_{N2}
	            \end{array}
	         \\
	        \begin{array}{c}
	          \tilde{q}^*_{11} \\
	          \vdots \\
	          \tilde{q}^*_{N1}
	        \end{array}
	         &  \begin{array}{c}
	              \mathbf{Q}^{T}
	            \end{array}
	\end{array}\right),
	\label{Ndet_SS}
\end{eqnarray}
where the new vectors $\mathbf{\tilde{q}}_{n}=(\tilde{q}_{n1},\tilde{q}_{n2})^{T}$ depend on the initial matrix $\mathbf{\Phi}_0$ and we discuss them separately for the multi-soliton and multi-breather cases. 
Note that the dressing procedure is local, i.e., $\psi_{n}$ and $\mathbf{\Phi}_{n}$ depend on $\psi_{n-1}$ and $\mathbf{\Phi}_{n-1}$ taken at the same point $(x,t)$.


\subsection{Multi-soliton and multi-breather solutions}
\label{Sec:Methods:B}

\begin{figure}[t]\centering
	\includegraphics[width=0.99\linewidth]{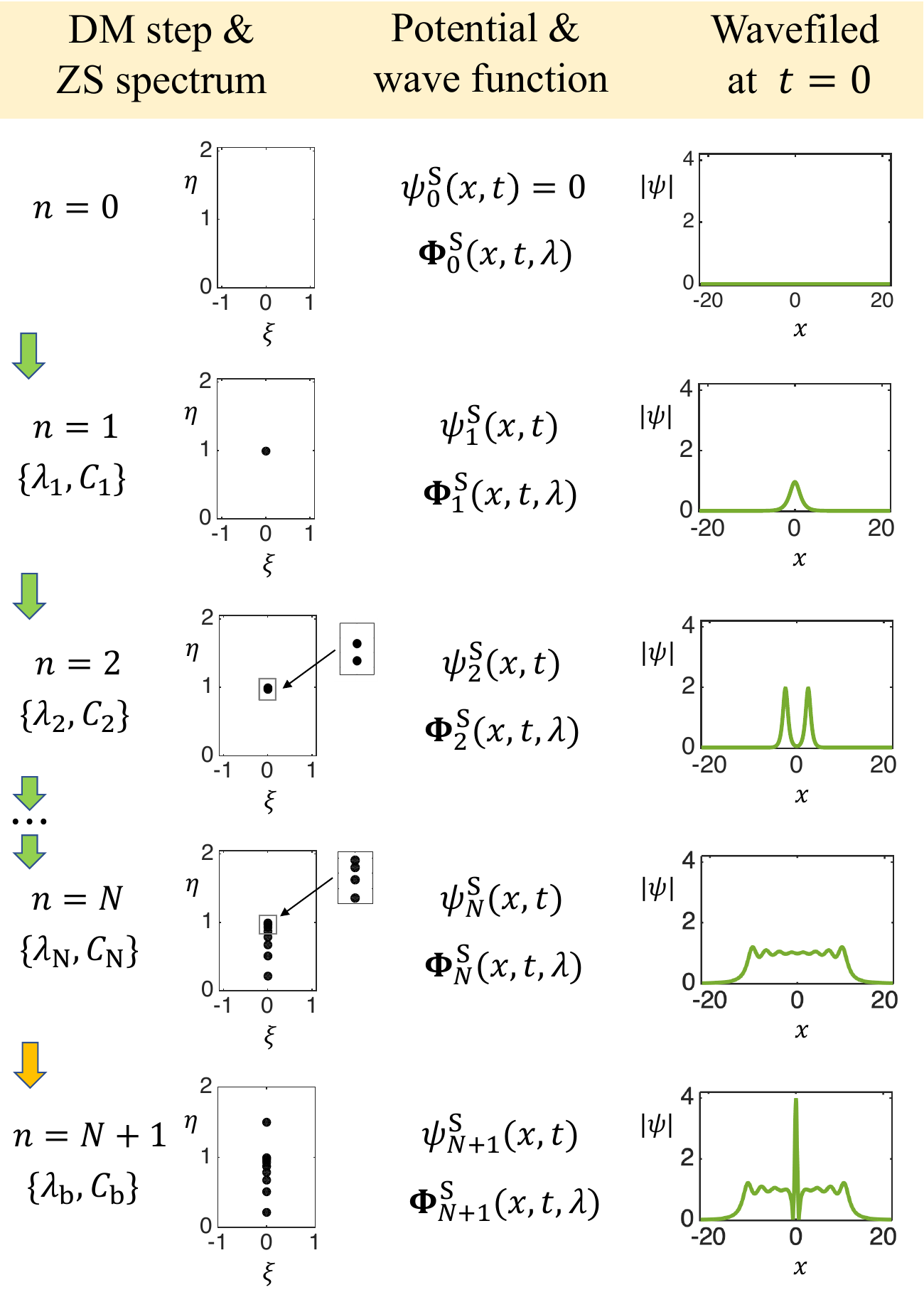}
	\caption{\small {\it (Color on-line)} 
            Schematic representation of the DM for constructing exact multi-soliton solutions. 
            The left column shows the step number $n$, the soliton $\{\lambda_{n},C_{n}\}$ used for dressing at this step, and the resulting eigenvalue spectrum of the ZS system~(\ref{ZSsystem1-eigenvalue}) in the upper half of the $\lambda$-plane. 
            The center column indicates the potential $\psi_{n}^{\mathrm{S}}(x,t)$ and wave function $\mathbf{\Phi}_{n}^{\mathrm{S}}(x,t,\lambda)$ constructed at this step, and the right column illustrates the potential $|\psi_{n}^{\mathrm{S}}(x)|$ at $t=0$. 
            Note that, for solitonic models of breathers, the neighboring eigenvalues $\lambda_{n-1}$ and $\lambda_{n}$ are very close to each other at small $n$; see the zoom of the ZS spectrum at steps $n=2$ and $N$. 
            Also note that for $n=N+1$ the soliton is designated as $\{\lambda_{\mathrm{b}},C_{\mathrm{b}}\}$ for subsequent comparison with breather solutions.
	}
	\label{fig:fig1}
\end{figure}

An exact $N$-soliton potential $\psi_{N}^{\mathrm{S}}$ is constructed from the zero background $\psi_{0}^{\mathrm{S}}(x,t)=0$ and the corresponding wave function
\begin{eqnarray}
	\mathbf{\Phi}_{0}^{\mathrm{S}}(x,t,\lambda) = \begin{pmatrix}\ e^{-i\lambda x - i \lambda^2 t} & 0 \\ 0 & e^{i\lambda x + i \lambda^2 t} \end{pmatrix},
	\label{Psi0}
\end{eqnarray}
with the outcome written as
\begin{eqnarray}
	\psi_{N}^{\mathrm{S}}(x,t) = -2i\frac{\mathrm{det} \mathbf{P}}{\mathrm{det} \mathbf{Q}},
	\label{NSsolution}
\end{eqnarray}
and the matrices $\mathbf{P}$ and $\mathbf{Q}$ defined via vectors
\begin{eqnarray}
	\mathbf{\tilde{q}}_{n}(x,t) = \left(\begin{array}{c}\tilde{q}_{n1} \\\tilde{q}_{n2}\end{array}\right) =
	\left(\begin{array}{c} C_n^{-1/2}\, e^{i\lambda_n x +  i \lambda^2 t} \\ C_n^{1/2}\, e^{-i\lambda_n x -  i \lambda^2 t} \end{array}\right).
	\label{tildeq-N-soliton}
\end{eqnarray}
Each step of the DM leads to the new potential $\psi_{n}^{\mathrm{S}}$ and new wave function $\mathbf{\Phi}_{n}^{\mathrm{S}}$, and consists in adding one soliton $\lambda_{n}$ to $\psi_{n-1}^{\mathrm{S}}$ and one pole at $\lambda=\lambda_{n}$ to $\mathbf{\Phi}_{n-1}^{\mathrm{S}}$ (we consider only the upper half of the $\lambda$-plane). 
This procedure is schematically illustrated in Fig.~\ref{fig:fig1}, where the last soliton is designated as $\{\lambda_{\mathrm{b}},C_{\mathrm{b}}\}$ for subsequent comparison with breather solutions. 
Note that, in Eq.~(\ref{tildeq-N-soliton}), the exponents in the right-hand side can be used to calculate the norming constants that will lead to the same dressed solution if the DM is performed at a different moment of time (that is, the time evolution of the norming constants),
\begin{eqnarray}
	C_{n}(t) = C_{n}(0)\,e^{-2i\lambda_{n}t}.
	\label{C_param_S-evolution}
\end{eqnarray}

If one parameterizes the norming constants $C_{n}$ via the soliton positions $x_{n}\in\mathbb{R}$ and phases $\theta_{n}\in\mathbb{R}$ at $t=0$ as
\begin{eqnarray}
	C_{n} &=& \exp\big[i\pi + 2i\lambda_{n}x_{n} + i\theta_{n}\big], \label{C_param_S}
\end{eqnarray}
then the one-soliton potential takes the following form,
\begin{eqnarray}
	\psi_{1}^{\mathrm{S}}(x,t) = a_{1} \frac{\exp\big[iv_{1}(x-x_{1}) + \frac{i(a_{1}^{2}-v_{1}^{2})t}{2} + i\theta_{1}\big]}{\cosh a_{1} \big(x-v_{1}t-x_1\big)},
	\label{1-SS}
\end{eqnarray}
where $a_{1}=2\eta_{1}$ and $v_{1}=-2\xi_{1}$ are the soliton amplitude and velocity. 
Note that parametrization~(\ref{C_param_S}) serves only for a convenient and physically intuitive representation of multi-soliton solutions, and that $x_{n}$ and $\theta_{n}$ coincide with the observed in the physical space position and phase of a soliton only for one-soliton potential. 
In presence of other solitons or dispersive waves, the observed position and phase may differ considerably from these parameters.

\begin{figure}[t]\centering
	\includegraphics[width=0.99\linewidth]{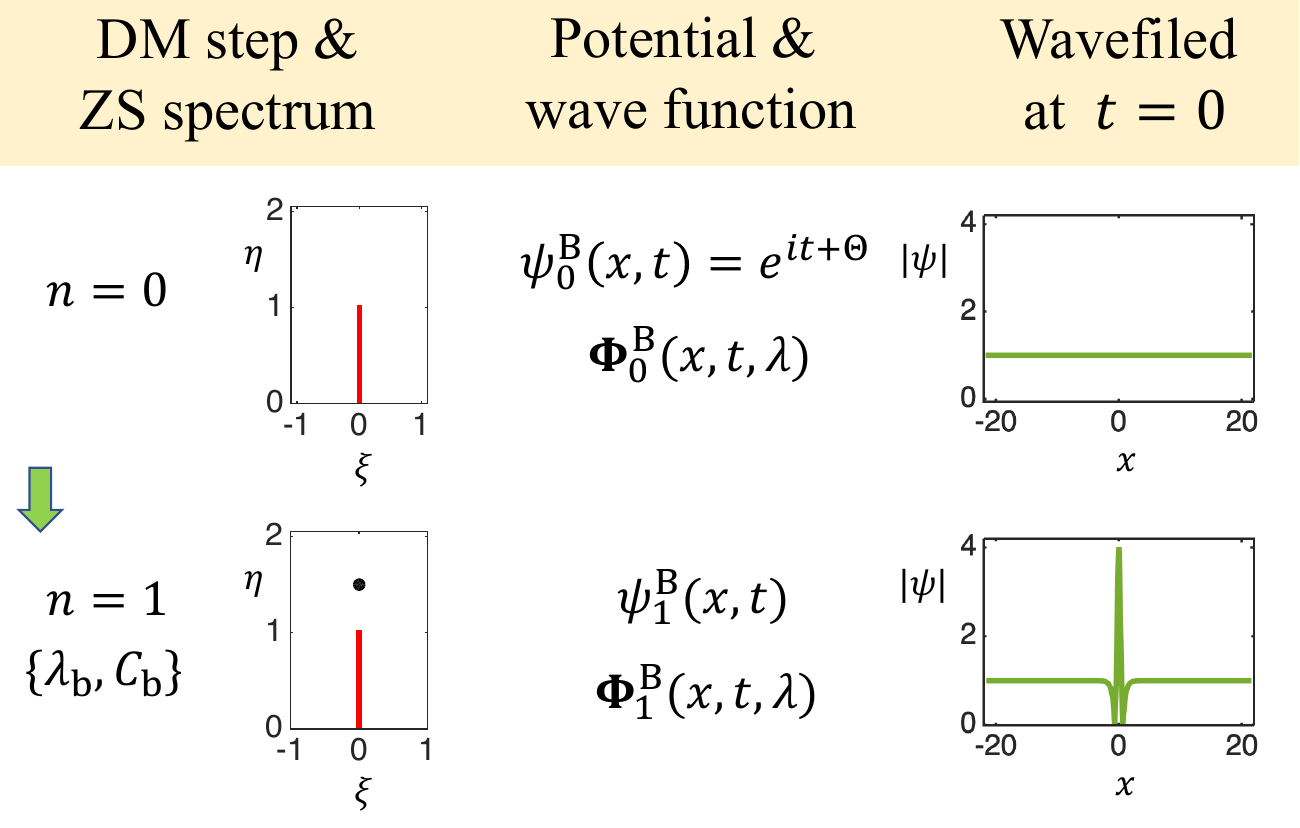}
	\caption{\small {\it (Color on-line)} 
            Same as in Fig.~\ref{fig:fig1} for constructing one-breather solutions. 
            The red line in the diagrams of the ZS spectrum shows the cut $[0,i]$, where wave functions $\mathbf{\Phi}_{n}^{\mathrm{B}}(x,t,\lambda)$ have non-analytic behavior.
	}
	\label{fig:fig2}
\end{figure}

An exact $N$-breather potential $\psi_{N}^{\mathrm{B}}$ is constructed from (i) the plane wave background $\psi_{0}^{\mathrm{B}}(x,t) = e^{it + i\Theta}$, where $\Theta\in\mathbb{R}$ is the initial phase, and (ii) the corresponding wave function,
\begin{eqnarray}
	&& \mathbf{\Phi}_{0}^{\mathrm{B}}(x,t,\lambda) = \begin{pmatrix}\ e^{\frac{i}{2}t-\phi}  &  p\, e^{i \Theta + \frac{i}{2}t+\phi}  \\  p\, e^{-i \Theta -\frac{i}{2}t-\phi}  & e^{-\frac{i}{2}t+\phi} \end{pmatrix},
	\label{Psi0cond}\\
	&& \phi = i\zeta(x + \lambda t), \quad p = i(\lambda-\zeta), \quad \zeta = \sqrt{1+\lambda^2}.
    \nonumber
\end{eqnarray}
Note that, as is usually done in the breather theory, for function $\zeta(\lambda)$ we use the branch cut $[-i,i]$, choosing the Riemann sheet such that $\mathrm{Im}\,\zeta>0$ at $\mathrm{Im}\,\lambda > 0$. 
This choice differs from the branch cut $i(-\infty,-1]\,\cup\, i[1,\infty)$ implied in software compilers. 
As we have noted in Section~\ref{Sec:Methods:A}, we consider only the upper half of the $\lambda$-plane, so that from now on we will refer to the branch cut as $[0,i]$. 
Also note that we consider the plane wave of unit amplitude, as other cases can be obtained with rescaling $t\to A^{2}t$, $x\to A\,x$ and $\psi\to \psi/A$. 

In the outcome of the DM, the plane wave $e^{it+i\Theta}$ can be extracted as a common factor, 
\begin{eqnarray}
    \psi_{N}^{\mathrm{B}}(x,t) =  e^{it+i\Theta} \bigg(1 - 2i\frac{\mathrm{det} \mathbf{P}}{\mathrm{det} \mathbf{Q}} \bigg),
    \label{NBsolution}
\end{eqnarray}
and the matrices $\mathbf{P}$ and $\mathbf{Q}$ are defined via vectors
\begin{eqnarray}
	&& \mathbf{\tilde{q}}_{n}(x,t) = 
    \\\nonumber
    && \left(\begin{array}{c} C_n^{-1/2} e^{i\frac{\Theta}{2}} e^{\phi_n} - C_n^{1/2} e^{-i\frac{\Theta}{2}} p_n e^{-\phi_n} \\ -C_n^{-1/2} e^{i\frac{\Theta}{2}} p_n e^{\phi_n} + C_n^{1/2} e^{-i\frac{\Theta}{2}} e^{-\phi_n} \end{array}\right), \label{qn_breathers}\\
	&& \phi_{n} = i\zeta_{n}(x + \lambda_{n}t), \,\,\, p_{n} = i(\lambda_{n}-\zeta_{n}), \,\,\, \zeta_{n} = \zeta(\lambda_{n}).
	\nonumber
\end{eqnarray}
The dressing procedure preserves the non-analytic behavior of the wave function on the branch cut, adding at each step one breather to the potential and one pole at $\lambda=\lambda_{n}$ to the wave function. 
A schematic representation of the DM for one-breather solution,
\begin{eqnarray}
	\psi_{1}^{\mathrm{B}}(x,t) =  e^{it+i\Theta} \bigg(1 - 4\eta_1 \frac{\tilde{q}_{11}^*\tilde{q}_{12}}{|\tilde{q}_{11}|^2+|\tilde{q}_{12}|^2} \bigg),
	\label{1Bsolution}
\end{eqnarray}
is shown in Fig.~\ref{fig:fig2}. 
Note that the plane wave and the resulting $N$-breather potentials are not localized and need to be considered by the quasi-periodic analogue of the IST called the finite-gap theory~\cite{novikov1984theory,osborne2010nonlinear}. 
However, the DM is based on a more general Lax pair representation and can still be used for constructing such solutions. 


\subsection{Fundamental breather solutions}
\label{Sec:Methods:C}

Depending on the positioning of the soliton eigenvalue relative to the branch cut $[0,i]$, a general one-breather potential represents one of the four different cases (i) $\lambda_{\mathrm{b}} = i$, (ii) $\lambda_{\mathrm{b}} = i\eta_{\mathrm{b}}$ with $\eta_{\mathrm{b}}<1$, (iii) $\lambda_{\mathrm{b}} = i\eta_{\mathrm{b}}$ with $\eta_{\mathrm{b}}>1$ and (iv) $\mathrm{Re}\,\lambda_{\mathrm{b}}\neq 0$, corresponding to the Peregrine, Akhmediev, Kuznetsov-Ma and Tajiri-Watanabe breathers respectively. 
Here we have changed the subscript to reflect that soliton $\lambda_{\mathrm{b}}$ is used for dressing the plane wave. 
In the following, we use different parametrizations of the norming constant $C_{\mathrm{b}}$ for different breathers. 
These parametrizations are needed to represent the corresponding solutions in a convenient and physically intuitive form, and also to place the breathers at a given position in space and time.

The Akhmediev breather is obtained by dressing a bare solution $\psi_{0}^{\mathrm{B}} = e^{it + i\Theta}$ with soliton $\lambda_{\mathrm{b}}=i\eta_{\mathrm{b}}$, $\eta_{\mathrm{b}}<1$. 
If one parametrizes the norming constant as
\begin{eqnarray}
	C_{\mathrm{b}}^{\mathrm{A}} = \exp\big[i(\pi+\Theta) + \Omega_{\mathrm{A}}t_{0} + i\theta_{0}\big], \label{C_param_AB}
\end{eqnarray}
then, this breather takes the following form,
\begin{eqnarray}\label{A-breather}
    &&\psi^{\mathrm{A}}(x,t) = e^{it+i\Theta+i\pi} \times\\
    &&\times\biggl(1 - \frac{2\kappa^2 \cosh[\Omega_{\mathrm{A}}(t-t_0)] + i\Omega_{\mathrm{A}} \sinh[\Omega_{\mathrm{A}}(t-t_0)]}
    {\cosh[\Omega_{\mathrm{A}}(t-t_0)] - \eta_{\mathrm{b}} \cos[2\kappa x+\theta_{0}]}\biggr), \nonumber
\end{eqnarray}
where $\kappa = \sqrt{1-\eta_{\mathrm{b}}^2}$ and $\Omega_{\mathrm{A}} = 2\eta_{\mathrm{b}}\kappa$, while $t_{0}\in\mathbb{R}$ and $\theta_{0}\in[0,2\pi)$ correspond to the time shift and space phase shift respectively.
The Akhmediev breather describes a spatially periodic perturbation with period $\pi/\kappa$ on the background of the plane wave $e^{it+i\Theta+i\pi}$; the perturbation becomes most pronounced at the time $t=t_{0}$ and vanishes within the characteristic time $\Omega_{A}^{-1}$ as $t\to\pm\infty$. 
Note that the additional phase shift $e^{i\pi}$ relative to the bare solution $\psi_{0}^{\mathrm{B}} = e^{it + i\Theta}$ appears at every new step of the dressing procedure; see below for details. 

The Kuznetsov-Ma breather is constructed by dressing with soliton $\lambda_{\mathrm{b}}=i\eta_{\mathrm{b}}$, $\eta_{\mathrm{b}}>1$. 
With parametrization
\begin{eqnarray}
	C_{\mathrm{b}}^{\mathrm{KM}} = \exp\big[i(\pi+\Theta) - 2\nu x_{0} + i\theta_{0}\big], \label{C_param_KMB}
\end{eqnarray}
this solution can be written as
\begin{eqnarray}\label{KM-breather}
    &&\psi^{\mathrm{KM}}(x,t) = e^{it+i\Theta+i\pi} \times\\
    &&\times\biggl(1 - \frac{2\nu^2 \cos[\Omega_{\mathrm{K}}t + \theta_{0}] + i\Omega_{\mathrm{K}} \sin[\Omega_{\mathrm{K}}t + \theta_{0}]}
    {\eta_{\mathrm{b}}\cosh[2\nu(x-x_0)] - \cos[\Omega_{\mathrm{K}}t + \theta_{0}]} \biggr), \nonumber
\end{eqnarray}
where $\nu = \sqrt{\eta_{\mathrm{b}}^2-1}$ and $\Omega_{\mathrm{K}} = 2\eta_{\mathrm{b}}\nu$, while $x_{0}\in\mathbb{R}$ and $\theta_{0}\in[0,2\pi)$ represent the space shift and time phase shift respectively.
This solution describes a standing localized perturbation of the plane wave $e^{it+i\Theta+i\pi}$; the perturbation has characteristic width $(2\nu)^{-1}$ in space and oscillates with period $2\pi/\Omega_{K}$ in time.

The Peregrine breather is obtained from the Kuznetsov-Ma solution in the limit $\epsilon = (\eta_{\mathrm{b}}-1) \to 0^{+}$, corresponding to the dressing with soliton $\lambda_{\mathrm{b}}=i$. 
Expanding Eq.~(\ref{KM-breather}) in Taylor series with respect to $\epsilon$ for $|x-x_{0}|\ll (8\epsilon)^{-1/2}$ and $|t|\ll (8\epsilon)^{-1/2}$, and using the following parametrization of the norming constant,
\begin{eqnarray}
    C_{\mathrm{b}}^{\mathrm{P}} = \exp\bigg[i(\pi+\Theta) - \sqrt{8\epsilon}\,\big(x_{0} + i\,t_{0}\big)\bigg], \label{C_param_PB}
\end{eqnarray}
we arrive to the Kuznetsov-Ma breather that behaves locally as the Peregrine breather,
\begin{eqnarray}\label{P-breather}
    &&\psi^{\mathrm{P}}(x,t) = e^{it+i\Theta+i\pi} \times\\
    &&\times\biggl(1 - \frac{4 (1 + 2 i [t-t_{0}])}{1 + 4 [x-x_{0}]^2 + 4 [t-t_{0}]^2} \biggr). \nonumber
\end{eqnarray}
Here $x_{0}\in\mathbb{R}$ and $t_{0}\in\mathbb{R}$ are the position and time shifts respectively. 
Note that formally, at $\epsilon=0$, there is only one allowed value of the norming constant $C_{\mathrm{b}} = e^{i\pi+i\Theta}$, which corresponds to the Peregrine breather~(\ref{P-breather}) for all possible $x_{0}$ and $t_{0}$. 
Any other value of $C_{\mathrm{b}}$ results in the plane wave solution $\psi^{\mathrm{P}} = e^{it+i\Theta+i\pi}$ describing the Peregrine breather located at infinity. 
The specific values of $x_{0}$ and $t_{0}$ are determined by how the norming constant~(\ref{C_param_PB}) tends to $e^{i\pi+i\Theta}$ in the limit $\epsilon\to 0^{+}$. 
The Peregrine breather describes a rational perturbation of the plane wave $e^{it+i\Theta+i\pi}$, which is localized both in space and in time and leads to the maximum amplitude $|\psi^{\mathrm{P}}|=3$ at the point $(x_{0},t_{0})$.

The Tajiri-Watanabe breather corresponds to a general case when the soliton eigenvalue $\lambda_{\mathrm{b}}$ has nonzero real and imaginary parts, so that we use Eq.~(\ref{1Bsolution}) without simplifications. 
This solution describes a coherent wave group on the background of the plane wave, which moves with a constant velocity $V_{\mathrm{TW}} = -\mathrm{Im}\,[\lambda_{\mathrm{b}}\zeta_{\mathrm{b}}]/\mathrm{Im}\,\zeta_{\mathrm{b}}$ in space and oscillates with a frequency $\Omega_{\mathrm{TW}} = 4\,\mathrm{Re}\,[\lambda_{\mathrm{b}}\zeta_{\mathrm{b}}] - 4\, V_{\mathrm{TW}}\, \mathrm{Re}\,\zeta_{\mathrm{b}}$ in time~\cite{xu2019breather}; here $\zeta_{\mathrm{b}} = \zeta(\lambda_{\mathrm{b}})$. 
The following parametrization of the norming constant,
\begin{eqnarray}
	C_{\mathrm{b}}^{\mathrm{TW}} = \exp\big[i(\pi+\Theta) + 2i\zeta_{\mathrm{b}} x_{0} + i\theta_{0}\big], \label{C_param_TWB}
\end{eqnarray}
allows one to find the position $x_{0}$ and phase $\theta_{0}$ of this breather at the initial time $t=0$.

In terms of evolution of the amplitude $|\psi|$, the Tajiri-Watanabe breather represents a localized perturbation on the background of the plane wave. 
However, it has different phases of this background to the left and to the right of itself, 
\begin{eqnarray}
	\lim_{x\to\pm\infty}\frac{\psi^{\mathrm{TW}}(x,t)}{e^{it+i\Theta}} = \exp{\bigg(\pm 2i\,\mathrm{arg}[\zeta_{\mathrm{b}} + \lambda_{\mathrm{b}}]\bigg)},
	\label{TW-phase-jump0}
\end{eqnarray}
meaning that the corresponding perturbation of the wavefield $\psi$ is not localized. 
Note that Eq.~(\ref{TW-phase-jump0}) is obtained by taking the limits $x\to\pm\infty$ in Eq.~(\ref{1Bsolution}) and is valid for any localized one-breather solution. 
This equation can be rewritten by using the Zhukovsky transform to the variables $\alpha\in [0,2\pi)$ and $R\in [0,\infty)$,
\begin{eqnarray}
    \lambda = \frac{i}{2}\bigg(R\,e^{i\alpha} + \frac{e^{-i\alpha}}{R}\bigg), \quad 
    \zeta = \frac{i}{2}\bigg(R\,e^{i\alpha} - \frac{e^{-i\alpha}}{R}\bigg),
    \label{ZhT}
\end{eqnarray}
see e.g.~\cite{gelash2014superregular} for the application of this transform in the theory of breathers. 
In the new variables, $\mathrm{arg}[\zeta_{\mathrm{b}} + \lambda_{\mathrm{b}}] = \alpha_{\mathrm{b}} + \pi/2$ and the phase shift by $\pi/2$ can be moved to the denominator of the left-hand side of Eq.~(\ref{TW-phase-jump0}), leading to
\begin{eqnarray}
	\lim_{x\to\pm\infty}\frac{\psi^{\mathrm{TW}}(x,t)}{e^{it+i\Theta+i\pi}} = \exp\big(\pm 2i\alpha_{\mathrm{b}}\big).
	\label{TW-phase-jump}
\end{eqnarray}

In the general case of an $N$-breather potential containing only the Peregrine, Kuznetsov-Ma and Tajiri-Watanabe breathers, the phase of the plane wave background at $x\to\pm\infty$ can be found as a multiplication of factors~(\ref{TW-phase-jump}), 
\begin{eqnarray}
	\lim_{x\to\pm\infty}\frac{\psi_{N}^{\mathrm{B}}(x,t)}{e^{it+i\Theta+iN\pi}} = \exp\bigg[\pm 2i\sum_{n=1}^{N}\alpha_{n}\bigg].
	\label{N-phase-jump}
\end{eqnarray}
In some cases, e.g., for a pair of super-regular breathers, one can achieve the same phases in the limits $x\to\pm\infty$, that corresponds to localized perturbation of the plane wave~\cite{gelash2018formation}. 
As follows from Eq.~(\ref{N-phase-jump}), each new breather changes the phase of the plane wave background by $\pi$. 

In Section~\ref{Sec:Results}, we compare one-breather solutions with our multi-soliton solutions for the initial phase $\Theta=\pi$ and norming constant $C_{\mathrm{b}}=1$. 
These two values correspond to the canonical form of the breathers living on the background of the plane wave $e^{it}$ when the position, time and phase shifts equal zero, $x_{0}=t_{0}=\theta_{0} = 0$. 


\subsection{Solitonic model of the plane wave solution}
\label{Sec:Methods:D}

In~\cite{gelash2021solitonic}, we have developed a solitonic model of the plane wave $\psi_{0}^{\mathrm{B}}=e^{it+i\Theta}$ in the form of an exact $N$-soliton solution ($N$-SS) converging asymptotically to $\psi_{0}^{\mathrm{B}}$ at large number of solitons $N$. 
This $N$-SS contains standing solitons with eigenvalues $\lambda_{n}\in (0, i)$, see also~\cite{manakov1973nonlinear,klaus2002purely, tsoy2003interaction, desaix2003eigenvalues},
\begin{eqnarray}
    && \bigg\{
    \{\lambda_n\} = \{ i\eta_n \} \,\,\, \big\rvert \,\,\, \tan\big(L\sqrt{1-\eta_{n}^2}\big) = -\frac{\sqrt{1-\eta_{n}^2}}{\eta_{n}},
    \nonumber\\
    && 0<\eta_n<1, \quad n=1,..,N\bigg\},
    \label{box_eigenvalues}
\end{eqnarray}
where $L = \pi(N-1/4)$ is the characteristic width of this solution in the physical space. 
For eigenvalues sorted in \textit{descending} order, $\eta_m < \eta_l$ for $m>l$, the norming constants take the following simple form,
\begin{equation}
	C_n = e^{i\pi n + i\Theta}, \label{DM-norming-constants-corrected}
\end{equation}
and correspond to solitons sitting at the coordinate origin $x_{n}=0$ with phases $\theta_{n}=\Theta + \pi(n-1)$, see Eq.~(\ref{C_param_S}). 
Note that, for $N\gg 1$, soliton eigenvalues can also be found within the semi-classical approximation~\cite{lewis1985semiclassical},
\begin{equation}\label{sc_box_eigenvalues}
	\lambda_{n}^{(sc)} = i\sqrt{1-\bigg[\frac{\pi(n-1/2)}{L}\bigg]^2}.
\end{equation}

\begin{figure}[t]\centering
	\includegraphics[width=0.99\linewidth]{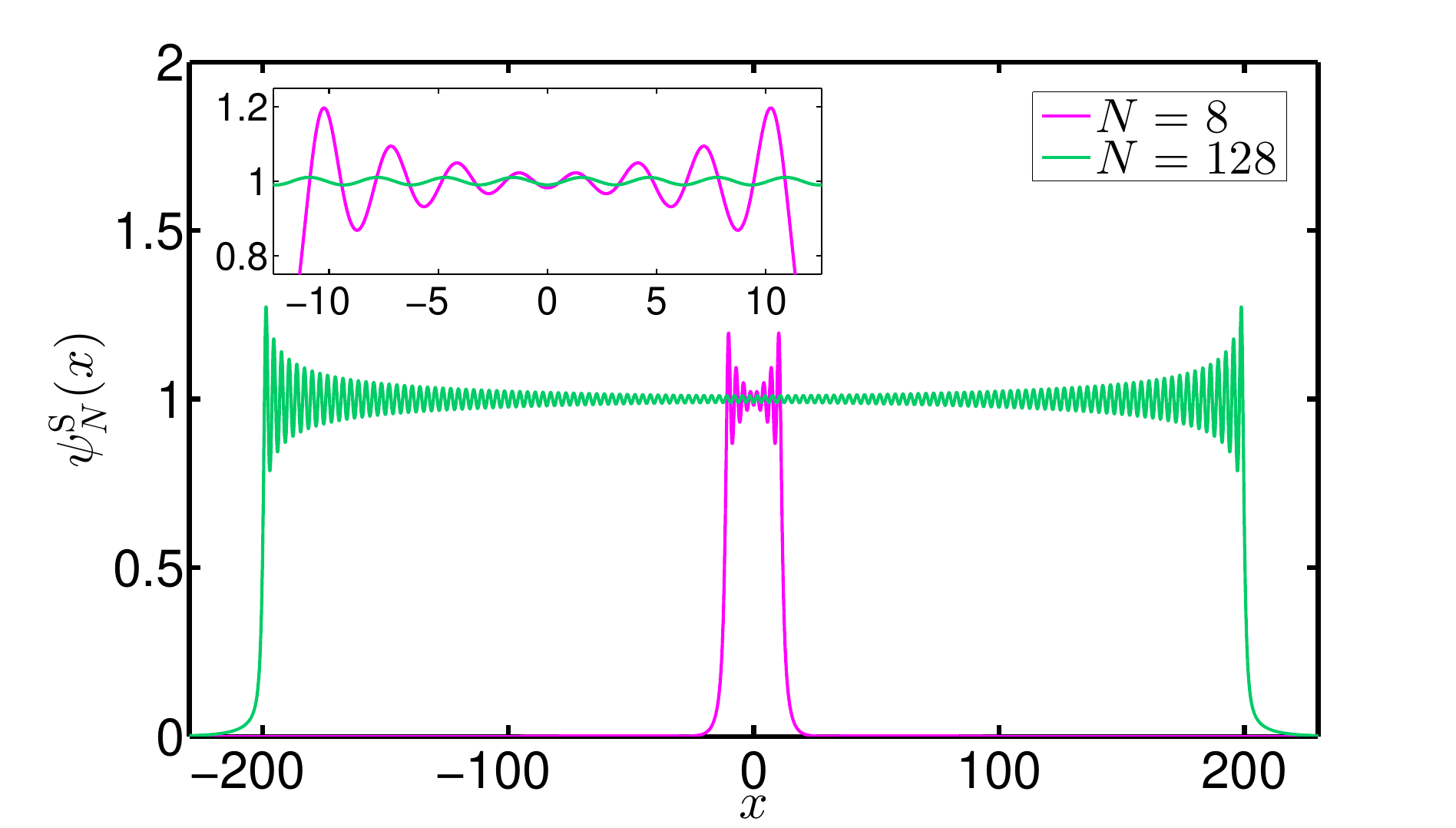}
	
	\caption{\small {\it (Color on-line)} 
		Solitonic model~(\ref{box_eigenvalues})-(\ref{DM-norming-constants-corrected}) of the plane wave $\psi_{0}^{\mathrm{B}}=e^{it+i\Theta}$ for $N=8$ (magenta) and $N=128$ (green), at the initial time $t=0$ and for the initial phase $\Theta=0$.
	}
	\label{fig:fig3}
\end{figure}

\begin{figure*}[t]\centering
    \includegraphics[width=0.99\linewidth]{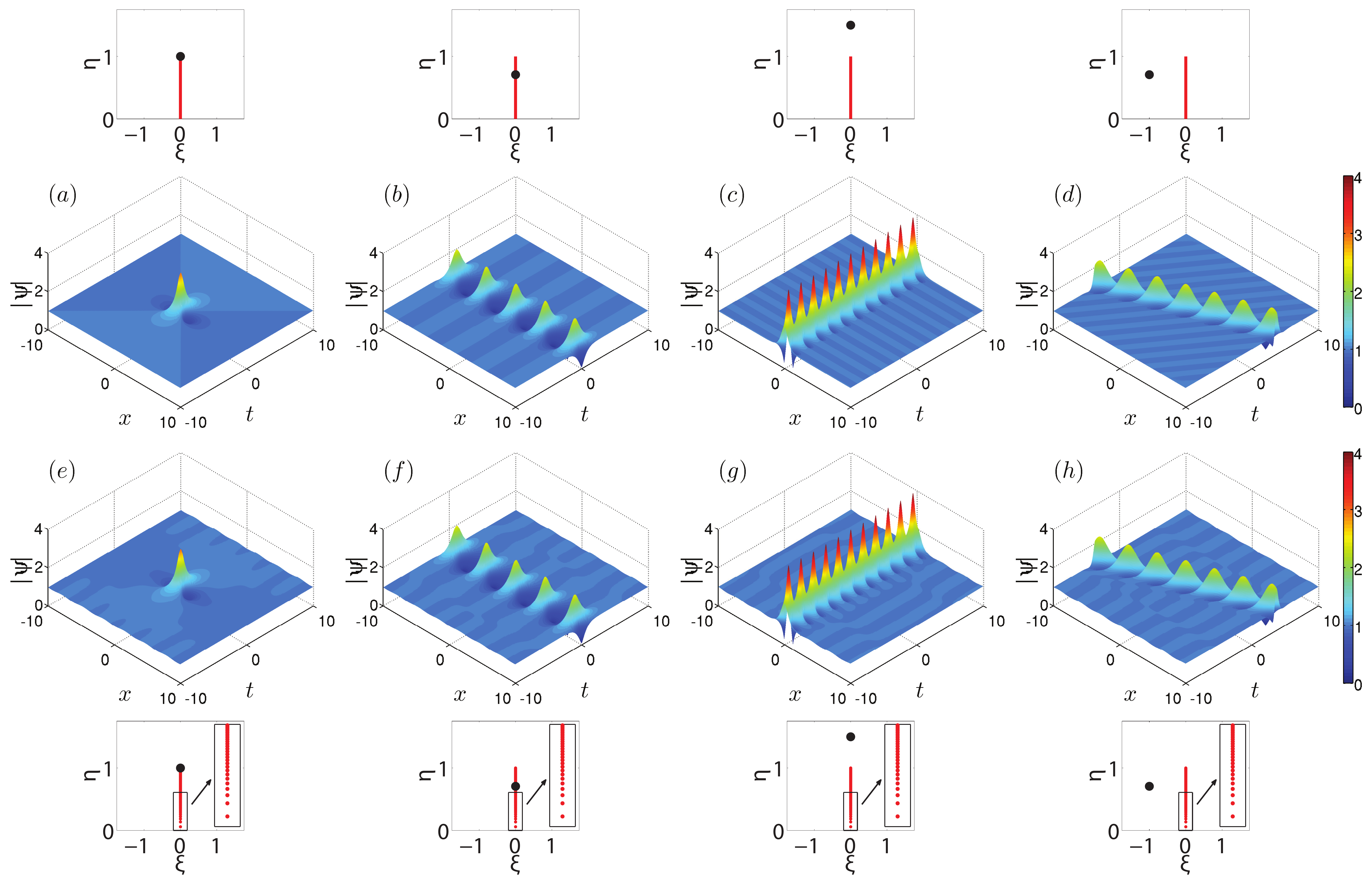}
    
	\caption{\small {\it (Color on-line)} 
		Space-time representation of (a) the Peregrine breather $\lambda_{\mathrm{b}}=i$, (b) the Akhmediev breather $\lambda_{\mathrm{b}}=i/\sqrt{2}$, (c) the Kuznetsov-Ma breather $\lambda_{\mathrm{b}}=1.5\,i$ and (d) the Tajiri-Watanabe breather $\lambda_{\mathrm{b}}=-1+i/\sqrt{2}$, and (e,f,g,h) their $129$-soliton models, respectively. 
        Above and below are shown the corresponding eigenvalue spectra in the upper half of the $\lambda$-plane: for exact breather solutions (a,b,c,d) -- a combination of the branch cut $[0,i]$ (red line) and soliton $\lambda_{\mathrm{b}}$ (black circle), while for $(N+1)$-soliton models (e,f,g,h) -- a combination of $N$ solitons modeling the plane wave~(\ref{box_eigenvalues}) (red dots) and soliton $\lambda_{\mathrm{b}}$ (black circle). 
        Note that, for the $128$-soliton model of the plane wave, the distance between neighbour eigenvalues is rather small, see the zoom in the corresponding figures. 
	}
	\label{fig:fig4}
\end{figure*}

\begin{figure*}[t]\centering
	\includegraphics[width=0.49\linewidth]{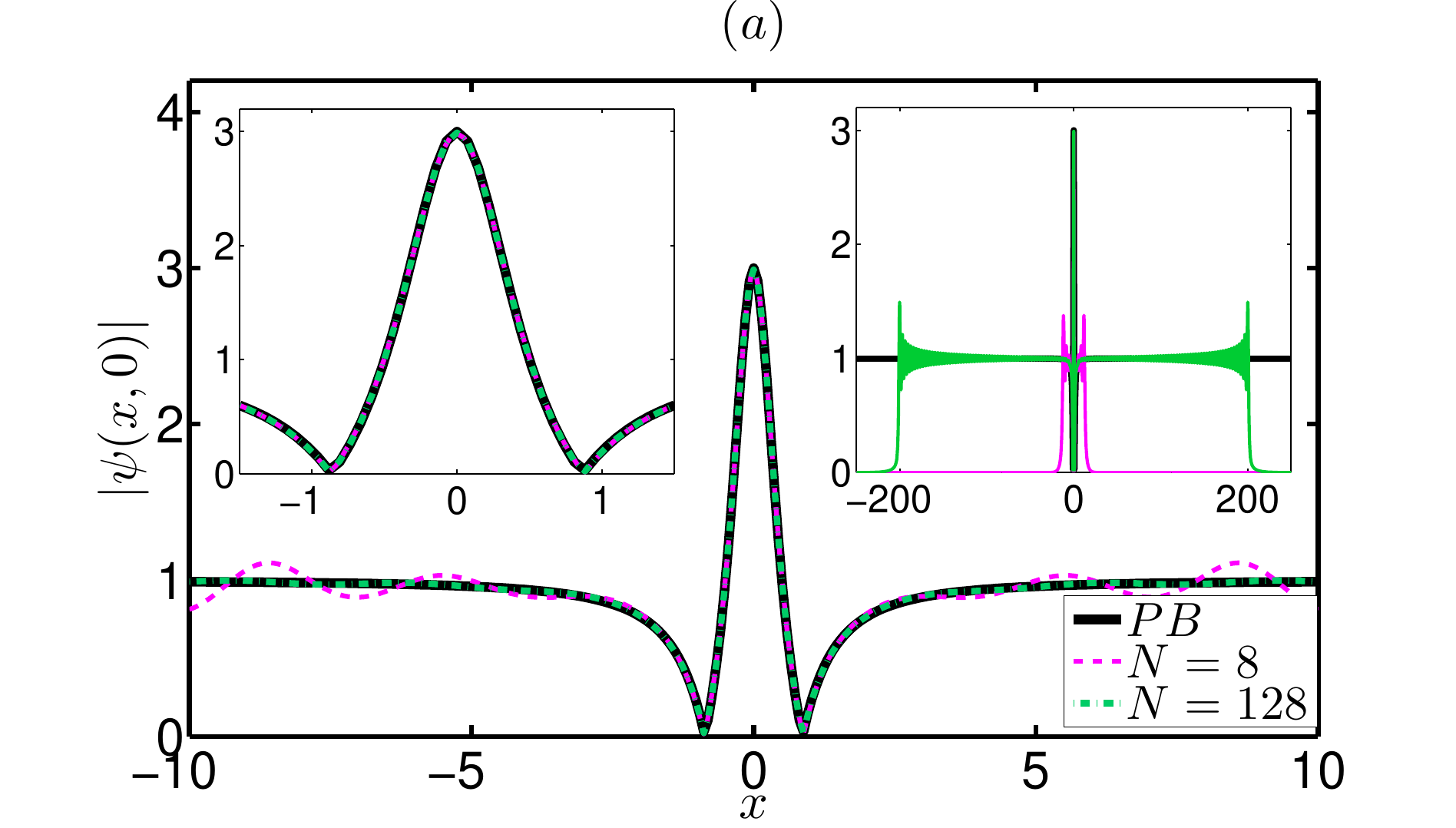}
	\includegraphics[width=0.49\linewidth]{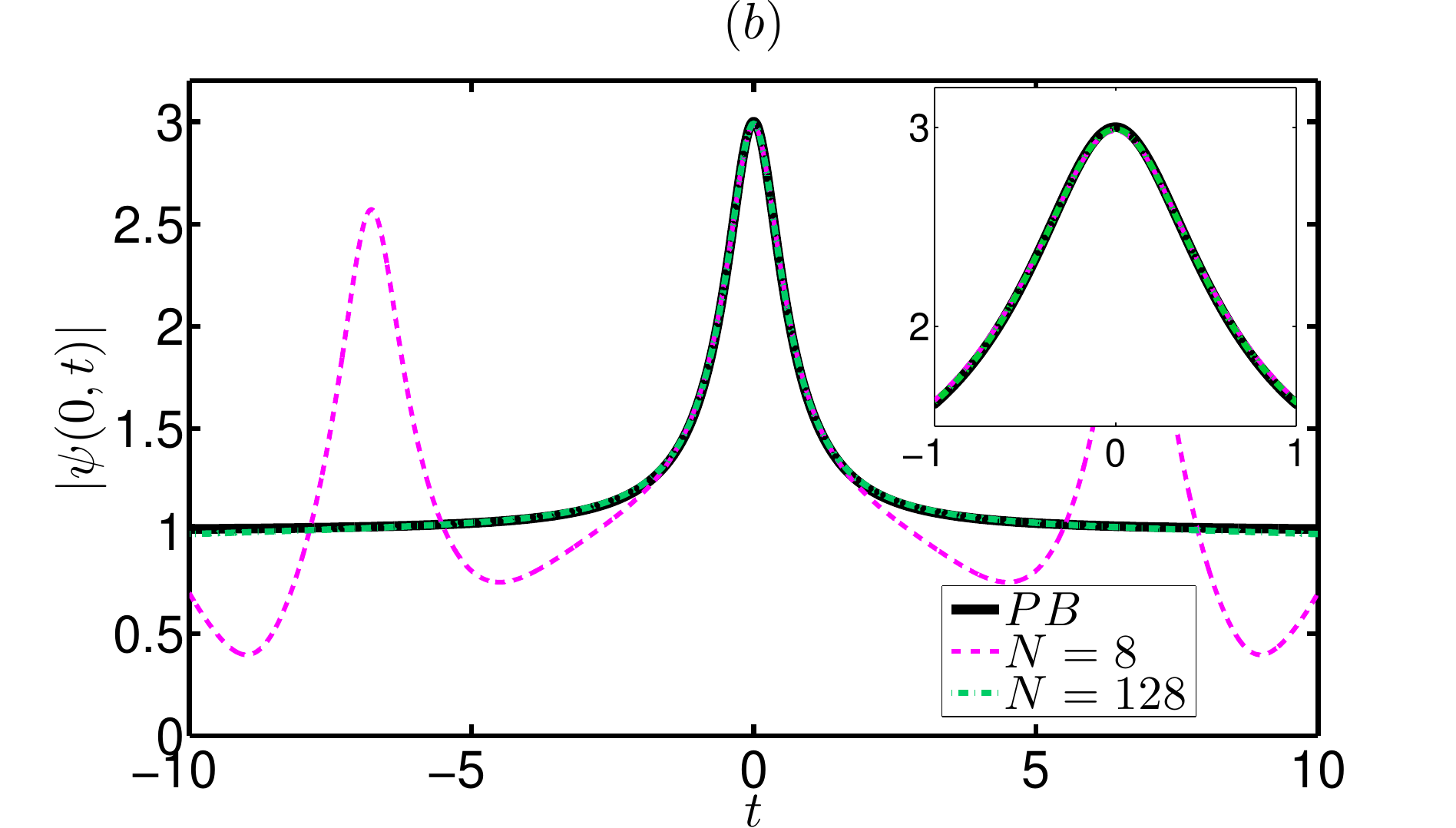}\\
	\includegraphics[width=0.49\linewidth]{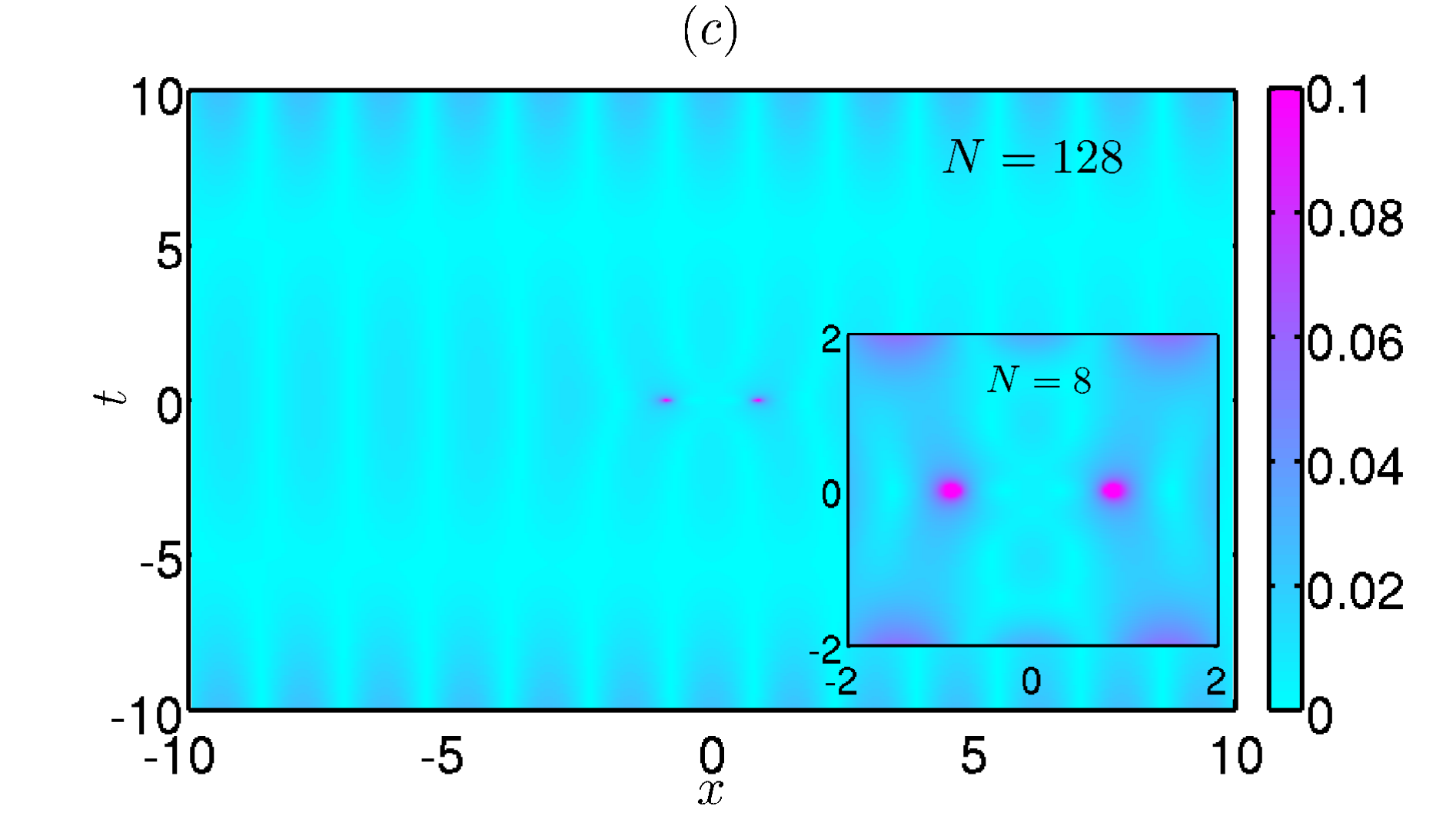}
	\includegraphics[width=0.49\linewidth]{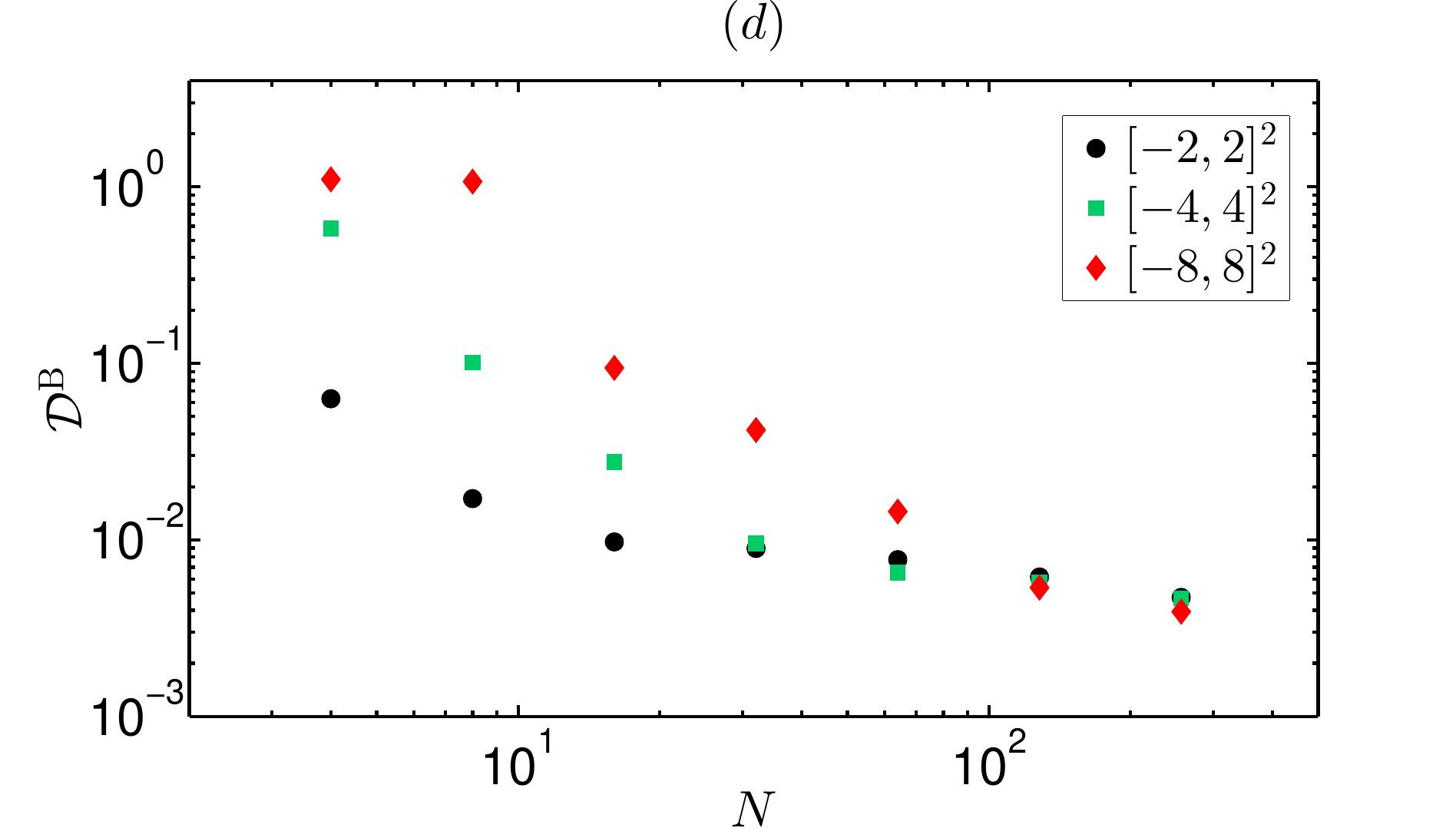}

	\caption{\small {\it (Color on-line)} 
		Peregrine breather (solid black) and its $(N+1)$-soliton model (dashed magenta for $N=8$ and dash-dot green for $N=128$): 
		(a) space profile $|\psi(x,t_{0})|$ at the time $t_{0}=0$ of the maximum elevation, 
		(b) time dependency $|\psi(x_{0},t)|$ at the maximum amplitude $x_{0}=0$, and 
		(c) relative deviation~(\ref{deviation-local}) in the $(x,t)$-plane for the $N=128$ model in the main panel and $N=8$ model in the inset (note the different scales). 
		Panel (d) shows the integral deviation~(\ref{deviation-integral}) between the Peregrine breather and its $(N+1)$-soliton model versus the number of solitons $N$ used for modeling the plane wave, for different regions $(x,t)\in[-\ell,\ell]\times[-\tau,\tau]$ where this deviation is calculated (note the double logarithmic scales). 
		The left inset in panel (a) and inset in panel (b) show zoom near the maximum amplitude, while the right inset in panel (a) demonstrates the solitonic model at large scale. 
		For better visualization, in panel (c), the deviations $d^{\mathrm{B}}(x,t)\ge 0.1$ are indicated with constant deep pink color. 
            Note that panels (a,b,c) are symmetric both in space and in time. 
	}
	\label{fig:fig5}
\end{figure*}

For $t=0$ and $\Theta=0$, the multi-soliton solution~(\ref{box_eigenvalues})-(\ref{DM-norming-constants-corrected}) has real-valued and symmetric in space wavefield, which models a part of a plane wave $\psi_{0}^{\mathrm{B}}(x)=1$ over the interval $|x|\lesssim L/2$; outside this interval, the wavefield is exponentially small. 
As shown in Fig.~\ref{fig:fig3} for the cases $N=8$ and $128$, in comparison with the plane wave, this $N$-SS contains residual oscillations due to the absence of nonlinear radiation. 
In~\cite{gelash2021solitonic}, we have demonstrated numerically that in a central region $|x|\le L_{s}/2$, which expands to the characteristic width $L$ with increasing $N$, these residual oscillations vanish by power law with $N$.

From a mathematical point of view, our construction of solitonic models of breathers assumes the continuity of the DM, meaning that if two potentials $\psi_{0}^{\mathrm{B}}$ and $\psi_{N}^{\mathrm{S}}$ are similar, then their dressing with the same soliton $\{\lambda_{\mathrm{b}}, C_{\mathrm{b}}\}$ leads to the similar results. 
Given that the dressing procedure is algebraic and local, this is possible when the corresponding wave functions $\mathbf{\Phi}_{0}^{\mathrm{B}}$ and $\mathbf{\Phi}_{N}^{\mathrm{S}}$ are close to each other in a broad region of space and time. 
In Appendix~\ref{Sec:App:C}, we verify numerically for some eigenvalues $\lambda_{\mathrm{b}}$ that this is indeed the case and the wave function $\mathbf{\Phi}_{N}^{\mathrm{S}}$ corresponding to the solitonic model~(\ref{box_eigenvalues})-(\ref{DM-norming-constants-corrected}) tends to the wave function $\mathbf{\Phi}_{0}^{\mathrm{B}}$ corresponding to the plane wave potential with increasing $N$.


\section{Solitonic models of breathers}
\label{Sec:Results}

In the dressing construction of one-breather solutions discussed in Section~\ref{Sec:Methods:C}, we replace the plane wave with its $N$-soliton model~(\ref{box_eigenvalues})-(\ref{DM-norming-constants-corrected}). 
In terms of the schematic illustration shown in Figs.~\ref{fig:fig1}-\ref{fig:fig2}, this corresponds to replacing the step $n=0$ in Fig.~\ref{fig:fig2} with the steps from $n=0$ to $N$ in Fig.~\ref{fig:fig1}, during which we build the solitonic model of the plane wave solution. 
Then, by dressing this solitonic model with the final soliton $\{\lambda_{\mathrm{b}}, C_{\mathrm{b}}\}$, we obtain $(N+1)$-soliton models $\psi_{N+1}^{\mathrm{P}}$, $\psi_{N+1}^{\mathrm{A}}$, $\psi_{N+1}^{\mathrm{KM}}$ and $\psi_{N+1}^{\mathrm{TW}}$ for the Peregrine, Akhmediev, Kuznetsov-Ma and Tajiri-Watanabe breathers respectively. 
In this Section, we compare these models numerically with the analytic solutions $\psi^{\mathrm{P}}$, $\psi^{\mathrm{A}}$, $\psi^{\mathrm{KM}}$ and $\psi^{\mathrm{TW}}$ for the same breathers. 
For definiteness, we use eigenvalues $\lambda_{\mathrm{b}}=i/\sqrt{2}$, $1.5\,i$ and $-1+i/\sqrt{2}$ for the Akhmediev, Kuznetsov-Ma and Tajiri-Watanabe breathers respectively; we have tried other values and came to the same results.

Note that the numerical calculation of multi-soliton wavefields containing a large number of solitons is a highly nontrivial problem due to the large number of arithmetic operations with exponentially small and large numbers. 
This problem has been solved only recently in~\cite{gelash2018strongly} with the help of the dressing method and high-precision arithmetics. 
In the present paper, we use the same scheme to construct multi-soliton solutions at the initial time $t=0$. 
To find these solutions at other times, we perform direct numerical simulation of the 1D-NLSE~(\ref{NLSE}) forward and backward in time using the Runge-Kutta fourth-order pseudospectral scheme on adaptive grid, see~\cite{agafontsev2015integrable}. 
An alternative approach would be to repeat the dressing construction at other times; however, our direct simulation provides the same results and takes much less computational resources. 

Figure~\ref{fig:fig4} shows the spatiotemporal representation of the analytic breather solutions $\psi^{\mathrm{P}}$, $\psi^{\mathrm{A}}$, $\psi^{\mathrm{KM}}$, $\psi^{\mathrm{TW}}$ (a,b,c,d) and our solitonic models $\psi_{N+1}^{\mathrm{P}}$, $\psi_{N+1}^{\mathrm{A}}$, $\psi_{N+1}^{\mathrm{KM}}$, $\psi_{N+1}^{\mathrm{TW}}$ for $N=128$ (e,f,g,h), together with the corresponding diagrams of the ZS spectrum. 
One can see a remarkable correspondence between breathers and our multi-soliton approximations over a wide region of space and time. 
In the following, we examine this correspondence in detail for the case of the Peregrine breather, leaving the other breathers to Appendix~\ref{Sec:App:A}. 

We start with the spatial profile of the Peregrine breather and its $(N+1)$-soliton model for $N=8$ and $N=128$, demonstrated in Fig.~\ref{fig:fig5}(a) at the time $t=0$ of the maximum elevation. 
As illustrated in the inset of the figure, the profile between the two roots $\psi^{\mathrm{P}}(x,0)=0$ is accurately reproduced already by the $9$-soliton solution ($9$-SS). 
The $129$-SS turns out to be practically indistinguishable from the breather over a wide space: in particular, the difference between the two solutions exceeds $0.1$ only closer to the edges of the $129$-SS, $|\psi_{129}^{\mathrm{P}}(x,0)-\psi^{\mathrm{P}}(x,0)|< 0.1$ at $|x|<185$. 
Comparing the time evolution of the solitonic model and the breather at the point $x=0$, see Fig.~\ref{fig:fig5}(b), we observe that the $9$-SS starts to deviate noticeably at $|t|\gtrsim 2$, while the $129$-SS model remains close to the breather much longer time, $|\psi_{129}^{\mathrm{P}}(0,t)-\psi^{\mathrm{P}}(0,t)|< 0.1$ for $|t|< 50$.

More generally, the deviation of a solitonic model $\psi_{N+1}^{\mathrm{B}}$ from the corresponding breather $\psi^{\mathrm{B}}$ can be measured locally in space and time as 
\begin{equation}
	d^{\mathrm{B}}(x,t) = \frac{|\psi_{N+1}^{\mathrm{B}} - \psi^{\mathrm{B}}|}{|\psi^{\mathrm{B}}|}. \label{deviation-local}
\end{equation}
Figure~\ref{fig:fig5}(c) shows this deviation for the $129$-soliton model in the region $(x,t)\in [-10,10]^{2}$ (main figure) and for the $9$-soliton model in the region $(x,t)\in [-2,2]^{2}$ (inset). 
For both models, deviation~(\ref{deviation-local}) remains well within $2$\% for most of the areas demonstrated in the figure, and the maximum relative deviation is observed near the roots $\psi^{\mathrm{P}}(x,t)=0$. 

As an integral measure reflecting these deviations, one can consider a quantity
\begin{equation}
	D^{\mathrm{B}} = \bigg[\frac{\int_{-\ell}^{\ell}dx\int_{-\tau}^{\tau}dt\,|\psi_{N+1}^{\mathrm{B}} - \psi^{\mathrm{B}}|^{2}}{\int_{-\ell}^{\ell}dx\int_{-\tau}^{\tau}dt\,|\psi^{\mathrm{B}}|^{2}}\bigg]^{1/2}, \label{deviation-integral}
\end{equation}
calculated over the region $(x,t)\in[-\ell,\ell]\times[-\tau,\tau]$. 
Table~\ref{tab:tab1} summarizes our results for $D^{\mathrm{B}}$ for all four one-breather solutions, demonstrating that the accuracy of our solitonic approximations is very high both in space and in time. 
Note that deviation~(\ref{deviation-integral}) depends non-trivially on the number of solitons $N$ used for modeling the plane wave background, as indicated in Fig.~\ref{fig:fig5}(d) for three different regions $[-\ell,\ell]\times[-\tau,\tau]$. 
In particular, as long as the characteristic width of the solitonic model is smaller than the comparison window, $L\lesssim 2\,\ell$, deviation~(\ref{deviation-integral}) stays of unity order; then, it decreases sharply with increasing $N$, and at larger $N$ the decrease continues according to a less steep law.

\setlength{\tabcolsep}{6pt}
\begin{table}[t]
	\caption{Integral deviations~(\ref{deviation-integral}) between breathers and their solitonic models, for the Peregrine ($\mathrm{PB}$) $\lambda_{\mathrm{b}}=i$, Akhmediev ($\mathrm{AB}$) $\lambda_{\mathrm{b}}=i/\sqrt{2}$, Kuznetsov-Ma ($\mathrm{KMB}$) $\lambda_{\mathrm{b}}=1.5\,i$ and Tajiri-Watanabe ($\mathrm{TWB}$) $\lambda_{\mathrm{b}}=-1+i/\sqrt{2}$ breathers.
	}
	
	\begin{center}
		\begin{tabular}{| c | c | c | c |}
			\hline
                \multirow{2}{*}{Breather}	& $N=8$,    & \multicolumn{2}{c}{$N=128$}\vline\\ \cline{3-4}
                                & $[-2,2]^{2}$ 			& $[-2,2]^{2}$         & $[-8,8]^{2}$           \\ \hline
			$\mathrm{PB}$ 	& $1.7\times 10^{-2}$ 	& $6.2\times 10^{-3}$  & $5.4\times 10^{-3}$    \\ \hline  
			$\mathrm{AB}$ 	& $1.8\times 10^{-2}$ 	& $7.9\times 10^{-3}$  & $5.4\times 10^{-3}$    \\ \hline
			$\mathrm{KMB}$	& $2.2\times 10^{-2}$ 	& $4.6\times 10^{-3}$  & $4.7\times 10^{-3}$    \\ \hline
			$\mathrm{TWB}$	& $2.0\times 10^{-2}$ 	& $6.7\times 10^{-3}$  & $5.1\times 10^{-3}$    \\ \hline
		\end{tabular}
	\end{center}
	\label{tab:tab1}
\end{table}

It is worth noting that for the solitonic models of the Peregrine and Kuznetsov-Ma breathers, the addition of a single standing soliton to the $N$-soliton model of the plane wave~(\ref{box_eigenvalues})-(\ref{DM-norming-constants-corrected}) perturbs the wavefield of the latter mainly at the center of the solution, $|x|\lesssim 1$. 
For the Akhmediev breather case, on the contrary, the dressing procedure significantly perturbs the wavefield everywhere in space. 
For instance, in Fig.~\ref{fig:fig4}(b,f), the Akhmediev breather $\lambda_{\mathrm{b}}=i/\sqrt{2}$ at the initial time $t=0$ reaches its maximum values $\max|\psi^{\mathrm{A}}|=\sqrt{2}+1$ on the periodic set of points $x = \pi\sqrt{2}\,m$, $m\in\mathbb{Z}$, and this behavior is accurately reproduced by our solitonic model $\psi_{N+1}^{\mathrm{A}}$ over its entire characteristic width $|x|\lesssim L/2$; see Appendix~\ref{Sec:App:A} for details. 
The similar situation takes place for the Tajiri-Watanabe breather, since the dressing procedure changes the complex phase of the plane wave background to the left and to the right from the breather, see Eq.~(\ref{TW-phase-jump}). 
These examples demonstrate the extreme sensitivity of multi-soliton wavefields to changes in the eigenvalue spectrum, when adding just one soliton to a solution, that already contains many more solitons, significantly perturbs the wavefield of the latter everywhere in space.

In Appendix~\ref{Sec:App:A}, we present a detailed numerical comparison of our solitonic models versus the analytic solutions for the Akhmediev, Kuznetsov-Ma and Tajiri-Watanabe breathers, and in Appendix~\ref{Sec:App:B} we provide the similar results for the higher-order rational and super-regular breathers, which represent specific examples of $2$- and $3$-breather solutions of the 1D-NLSE.


\section{Conclusions and discussion}
\label{Sec:Conclusions}

We have presented a universal method for constructing exact multi-soliton solutions that approximate breather dynamics with a very high accuracy over a wide region of space and time. 
This method consists in replacing the plane wave background in the dressing construction of the breathers with solitonic model of this background~\cite{gelash2021solitonic} and then in repeating the dressing procedure with exactly the same parameters as used for the construction of breathers. 
The resulting solitonic models approximate breathers locally very well already for $N\sim 10$ solitons, while for $N\sim 100$ solitons they are practically indistinguishable from breathers over a wide region of space and time. 
We think that our method can be generalized straightforwardly to multi-breather solutions, breathers on a non-trivial background (e.g., cnoidal waves) and other integrable systems including vector breathers~\cite{kedziora2014rogue,baronio2014vector,chen2019rogue,xu2020observation,che2022nondegenerate,gelash2023vector,hoefer2023kdv}, making it possible to model the rich dynamics of complex breather interactions~\cite{dubard2013multi,frisquet2013collision,kedziora2013classifying,gelash2018formation,conforti2018auto,xu2019breather,gelash2022management} and the behaviour of breathers in nearly integrable systems~\cite{kimmoun2016modulation,coppini2020effect}. 
We assume that for such a generalization one will only need to find the solitonic model for the breather background. 
From a mathematical point of view, our method assumes the continuity of the dressing procedure when the same dressing of two similar solutions leads to the similar results. 
Finding conditions when this continuity is satisfied is an interesting problem for future studies. 

We believe that our results have several implications for studies of rogue waves and, more generally, nonlinear waves. 
First, as we have noted in the Introduction, the theory of rogue waves needs to be supplemented with localized solutions which evolve locally as breathers. 
Here we have presented the solitonic ``skeletons'' of such solutions, which (i) can be made of arbitrary characteristic width $L\propto N$ and (ii) become practically indistinguishable from breathers as $L$ increases. 
Note that the time evolution of our solitonic models is known exactly through the evolution of the norming constants~(\ref{C_param_S-evolution}).

\begin{figure}[t]\centering
	\includegraphics[width=0.99\linewidth]{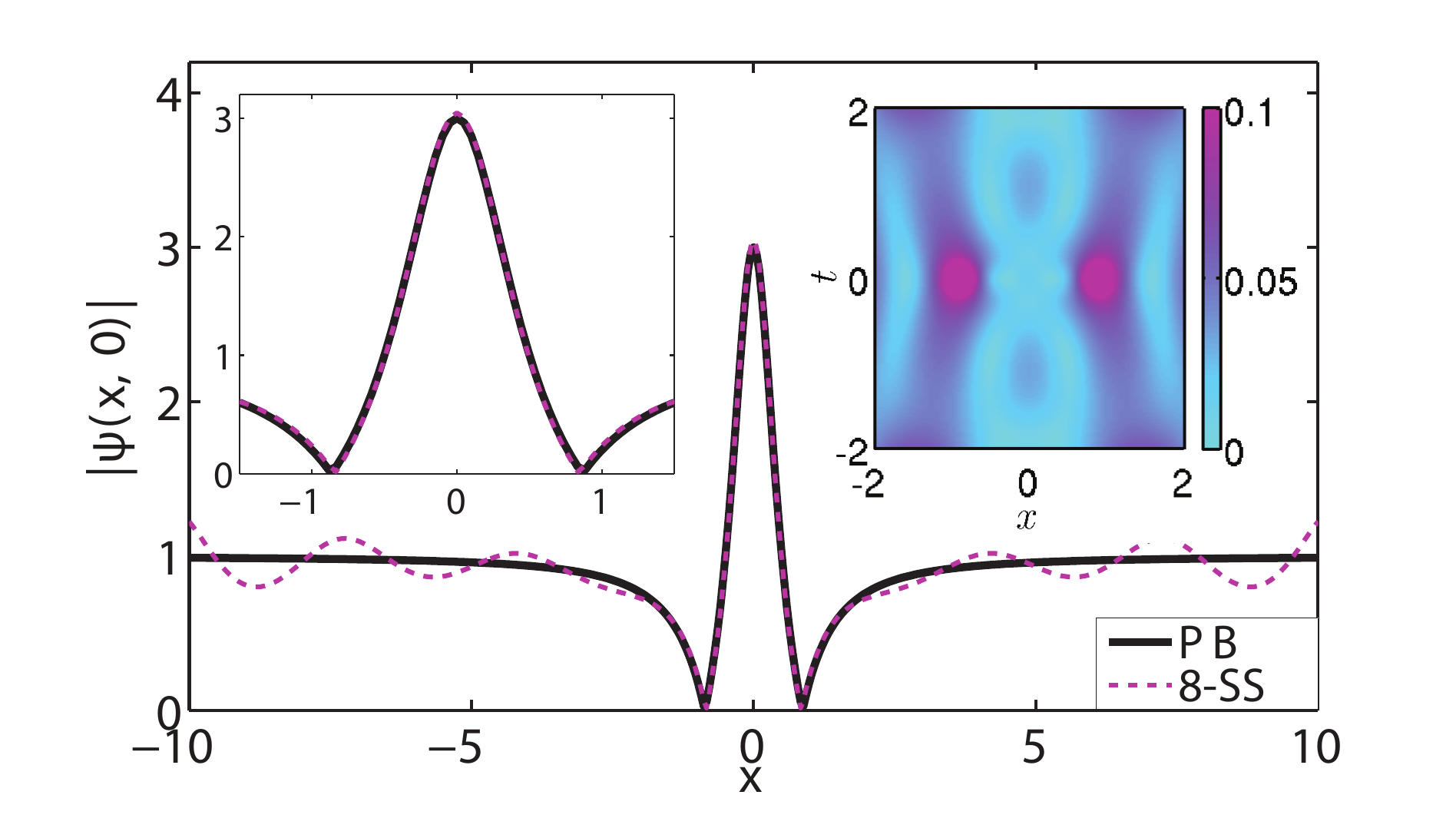}

	\caption{\small {\it (Color on-line)} 
		Peregrine breather (solid black) and its $8$-soliton model (dashed magenta) constructed from the $8$-soliton model of the plane wave with two modifications from the procedure discussed in Section~\ref{Sec:Results}. 
        First, we have used the semi-classical eigenvalues~(\ref{sc_box_eigenvalues}) instead of the exact values~(\ref{box_eigenvalues}). 
        Second, we have changed the norming constants of the $8$-soliton model of the plane wave to $C_{n}=1$, $1$, $-1$, $1$, ..., $-1$, $1$, for $n=1,...,8$, \textit{without the subsequent dressing procedure}. 
		The main panel shows the space profile $|\psi(x,t_{0})|$ at the time $t_{0}=0$ of the maximum elevation, the left inset demonstrates zoom near the maximum elevation, and the right inset illustrates the relative deviation~(\ref{deviation-local}) in the $(x,t)$-plane (deviations $d^{\mathrm{B}}(x,t)\ge 0.1$ are indicated with constant deep pink color). 
		The integral deviation~(\ref{deviation-integral}) in the region $(x,t)\in[-2,2]^{2}$ equals $3.0\times 10^{-2}$, that is about twice larger than for the $9$-soliton model in Fig.~\ref{fig:fig5}. 
	}
	\label{fig:fig6}
\end{figure}

Secondly, at large $N$, our solitonic models contain many objects and effectively represent \textit{soliton gas} models. 
In the present paper, we have found the parameters of these models to approximate classical breather solutions. 
However, both in numerical simulations and in real-world systems, rogue waves do not appear in such a ``pure'' form as classical breathers demonstrate, but instead they are surrounded by comparable chaotic perturbations of the wavefield. 
These perturbations cannot be described with the existing breather models which contain only a few breathers. 
Moreover, the existing breather models are deterministic and imply certain scenarios of rogue wave appearance, such as in the beginning of nonlinear stage of modulational instability (MI)~\cite{benjamin1967disintegration,zakharov1968theory} for the Akhmediev, Peregrine, higher-order rational and super-regular breathers~\cite{akhmediev1986modulation,shrira2010makes,akhmediev2009rogue,gelash2014superregular}. 
Also, they describe rogue waves that appear either once or periodically in time. 
However, in numerical simulations, breather-like rogue waves appear randomly from time to time, and this happens in various scenarios, for instance, long after the MI is fully developed~\cite{agafontsev2015integrable,agafontsev2016integrable,agafontsev2021rogue}, or in the long-time evolution from partially coherent waves~\cite{agafontsev2021extreme}, or in nearly integrable systems~\cite{agafontsev2015intermittency}. 
The existing breather models are incapable to capture this behavior, but it can be adequately described by soliton gas models. 

Indeed, the spontaneous appearance of rogue waves has already been observed in soliton gas~\cite{agafontsev2021rogue}; moreover, the dynamical and statistical properties of these rogue waves matched with those in the long-time evolution from the spontaneous modulational instability~\cite{agafontsev2015integrable}. 
Note that according to very recent studies~\cite{agafontsev2020growing,agafontsev2023bound} of nearly integrable systems with weak pumping, soliton gas can be a very common object in nature, as it becomes dominant already at weak nonlinearity. 
In soliton gas~\cite{agafontsev2021rogue}, the eigenvalues $\lambda_{n}$ were the same as for the solitonic model of the plane wave~(\ref{box_eigenvalues})-(\ref{DM-norming-constants-corrected}), the soliton positions $x_{n}$ were distributed within a small interval around the coordinate origin, and the phases $\theta_{n}$ were random. 
In the solitonic models of the present paper, the eigenvalues are distributed differently (as we add solitons to the solitonic model of the plane wave), but this difference does not lead to a significant effect for certain breathers. 

In particular, in Fig.~\ref{fig:fig6} we have constructed an $8$-soliton model of the Peregrine breather with two modifications from the discussed above procedure. 
First, we have used the semi-classical eigenvalues~(\ref{sc_box_eigenvalues}) instead of the exact values~(\ref{box_eigenvalues}). 
Second, we have started from the $8$-soliton model of the plane wave and simply changed its norming constants to $C_{n}=1$, $1$, $-1$, $1$, ..., $-1$, $1$ (to the same ones that would be for the $8$-soliton model of the Peregrine breather in Section~\ref{Sec:Results}), \textit{without the subsequent dressing procedure}. 
Thus, the only difference between this modified construction and what was discussed earlier is the inaccuracy in the eigenvalues. 
Despite this inaccuracy, the constructed solitonic model still shows a very good correspondence with the breather. 
We have repeated this construction for the Akhmediev and higher-order rational breathers using both the semi-classical and exact eigenvalues, and came to the similar results. 

Note that according to Eqs.~(\ref{C_param_S-evolution})-(\ref{C_param_S}), the soliton positions and phases evolve linearly with time,
\begin{eqnarray}
	x_{n}(t) &=& x_{n0} - 2\xi_{n}t, \label{evolution-xn}\\
	\theta_{n}(t) &=& \theta_{n0} + 2\left(\xi_{n}^{2} + \eta_{n}^{2}\right)t,\label{evolution-thetan}
\end{eqnarray}
where $x_{n0}$ and $\theta_{n0}$ are the positions and phases at $t=0$. 
In the general case, the frequencies $2\left(\xi_{n}^{2} + \eta_{n}^{2}\right)$ are incommensurable and may lead to spontaneous synchronization of soliton norming constants and the appearance of rogue wave. 
For instance, we can let the solitonic model of the plane wave~(\ref{box_eigenvalues})-(\ref{DM-norming-constants-corrected}) evolve, and at various moments of time it will turn into a very good approximation of the Akhmediev, Peregrine and high-order rational breathers.

We believe that the more general synchronization conditions for the soliton norming constants can be found, that will correspond to the emergence of rogue waves surrounded by chaotic perturbations of the wavefield. 
These rogue waves will appear spontaneously from time to time due to the spontaneous synchronization of soliton norming constants from one synchronization condition to another during the evolution in time. 
Finding these conditions represents a challenging problem for future studies.

Finally, our solitonic models may prove useful in explaining the process of soliton fission in nonintegrable systems described at leading order with the focusing 1D-NLSE. 
Indeed, if the initial wavefield can be approximated with an exact multi-soliton solution, in which most of the solitons are in a bound state, then the influence of nonintegrable perturbations is expected to gradually destroy this state by changing velocities differently for different solitons, thus leading to the fission. 
Note that soliton fission plays an important role in the supercontinuum generation~\cite{dudley2006supercontinuum} and formation of optical rogue waves~\cite{solli2007optical}; very recently it has been observed developing from the Peregrine and higher-order rational breathers~\cite{chowdury2023rogue,chowdury2022higher}. 


\begin{center}
\textbf{Acknowledgements}
\end{center}

The authors thank Gennady El for fruitful discussions. 
This work has been partially supported by the Agence Nationale de la Recherche through the LABEX CEMPI project (ANR-11-LABX-0007), the SOGOOD project (ANR-21-CE30-0061), the Ministry of Higher Education and Research, Hauts de France council and European Regional Development Fund (ERDF) through the Nord-Pas de Calais Regional Research Council and the European Regional Development Fund (ERDF) through the Contrat de Projets Etat-R\'egion (CPER Wavetech). 
The work on DM construction reported in Section~\ref{Sec:Methods} was supported by the Russian Science Foundation (Grant 19-72-30028 to DA and AG). 
The work of DA on construction of solitonic models of breathers reported in Section~\ref{Sec:Results} was supported by the state assignment of IO RAS, Grant FMWE-2021-0003. 
DA wishes to thank the Isaac Newton Institute and London Mathematical Society for the financial support on the Solidarity Programme, and also the Department of Mathematics, Physics and Electrical Engineering at Northumbria University for hospitality. 
The work of AG was funded by the European Union's Horizon 2020 research and innovation programme under the Marie Skłodowska-Curie grant No. 101033047. 
Simulations were performed at the Novosibirsk Supercomputer Center (NSU).


\appendix

\section{Solitonic models of one-breather solutions}
\label{Sec:App:A}

In Section~\ref{Sec:Results}, we have presented a detailed numerical comparison between the Peregrine breather $\psi^{\mathrm{P}}$ and its $(N+1)$-soliton model $\psi^{\mathrm{P}}_{N+1}$. 
Here we provide the similar results for the Akhmediev $\psi^{\mathrm{A}}$, Kuznetsov-Ma $\psi^{\mathrm{KM}}$ and Tajiri-Watanabe $\psi^{\mathrm{TW}}$ breathers versus their solitonic models $\psi^{\mathrm{A}}_{N+1}$, $\psi^{\mathrm{KM}}_{N+1}$ and $\psi^{\mathrm{TW}}_{N+1}$ respectively. 
We use parameters $\Theta=\pi$ and $C_{\mathrm{b}}=1$ for the dressing construction and set the eigenvalues to $\lambda_{\mathrm{b}}=i/\sqrt{2}$, $1.5\,i$ and $-1+i/\sqrt{2}$ respectively; we have tried other values and came to the same results.

\begin{figure*}[t]\centering
	\includegraphics[width=0.32\linewidth]{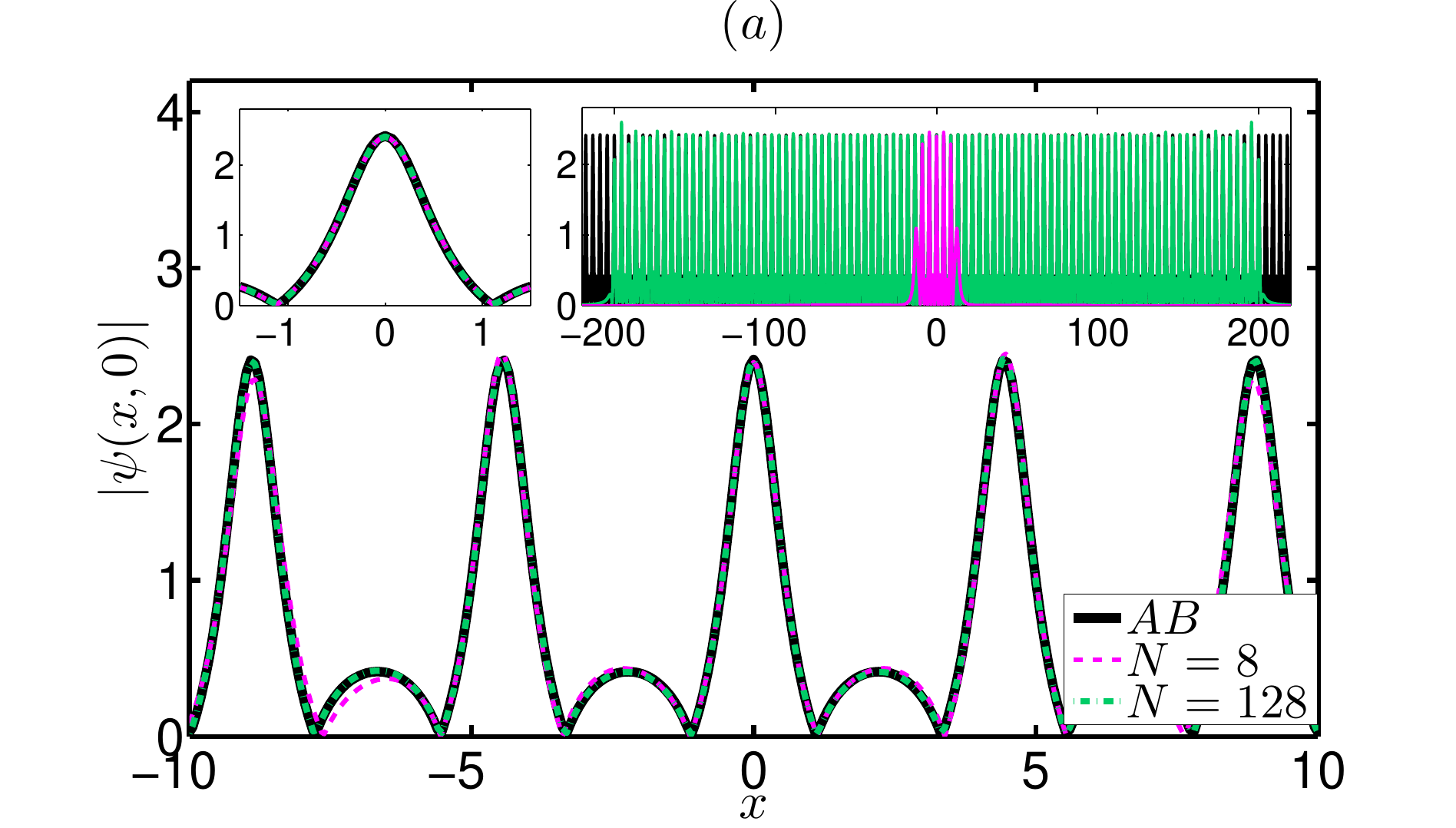}
	\includegraphics[width=0.32\linewidth]{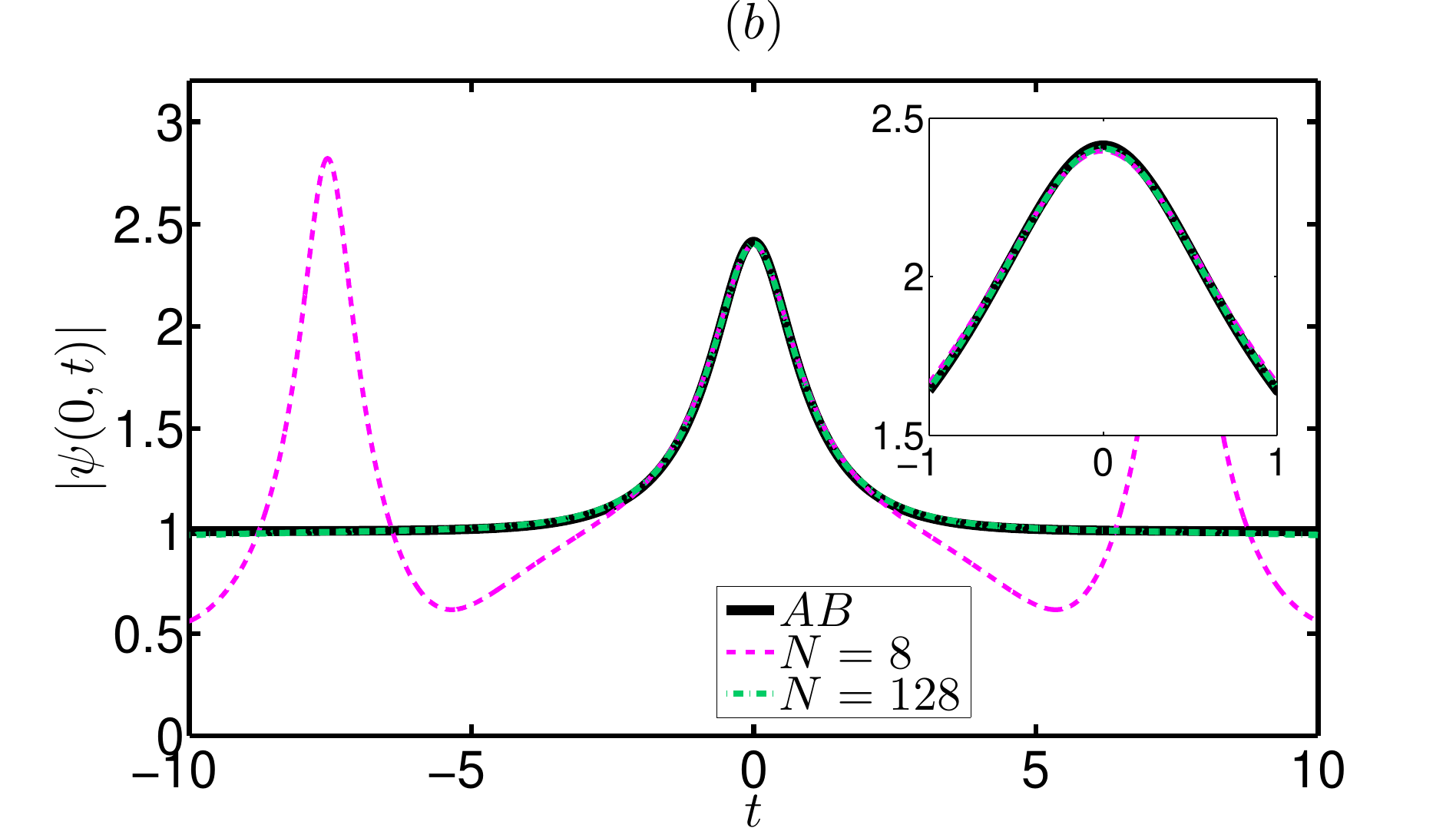}
	\includegraphics[width=0.32\linewidth]{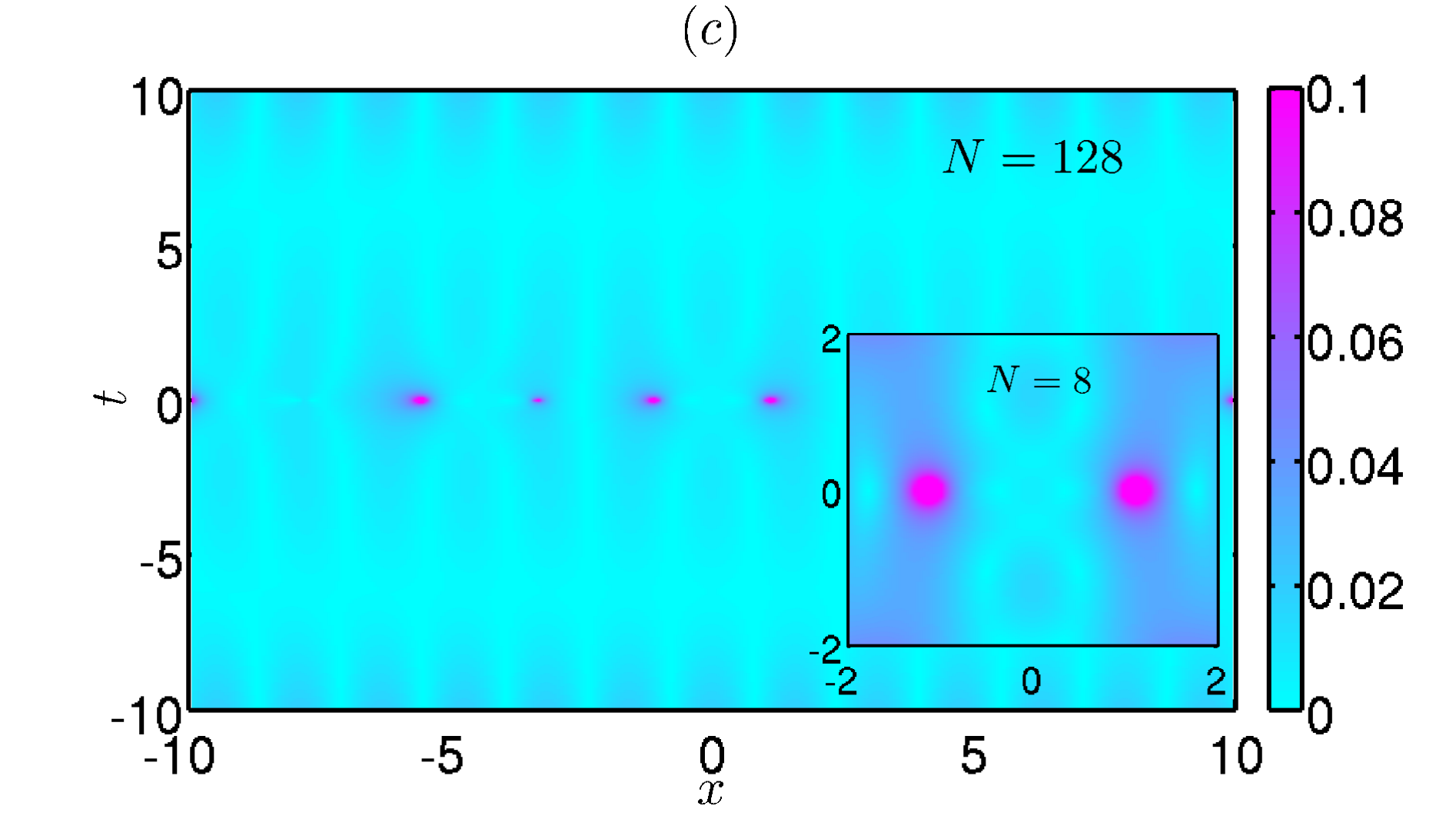}\\
	\includegraphics[width=0.32\linewidth]{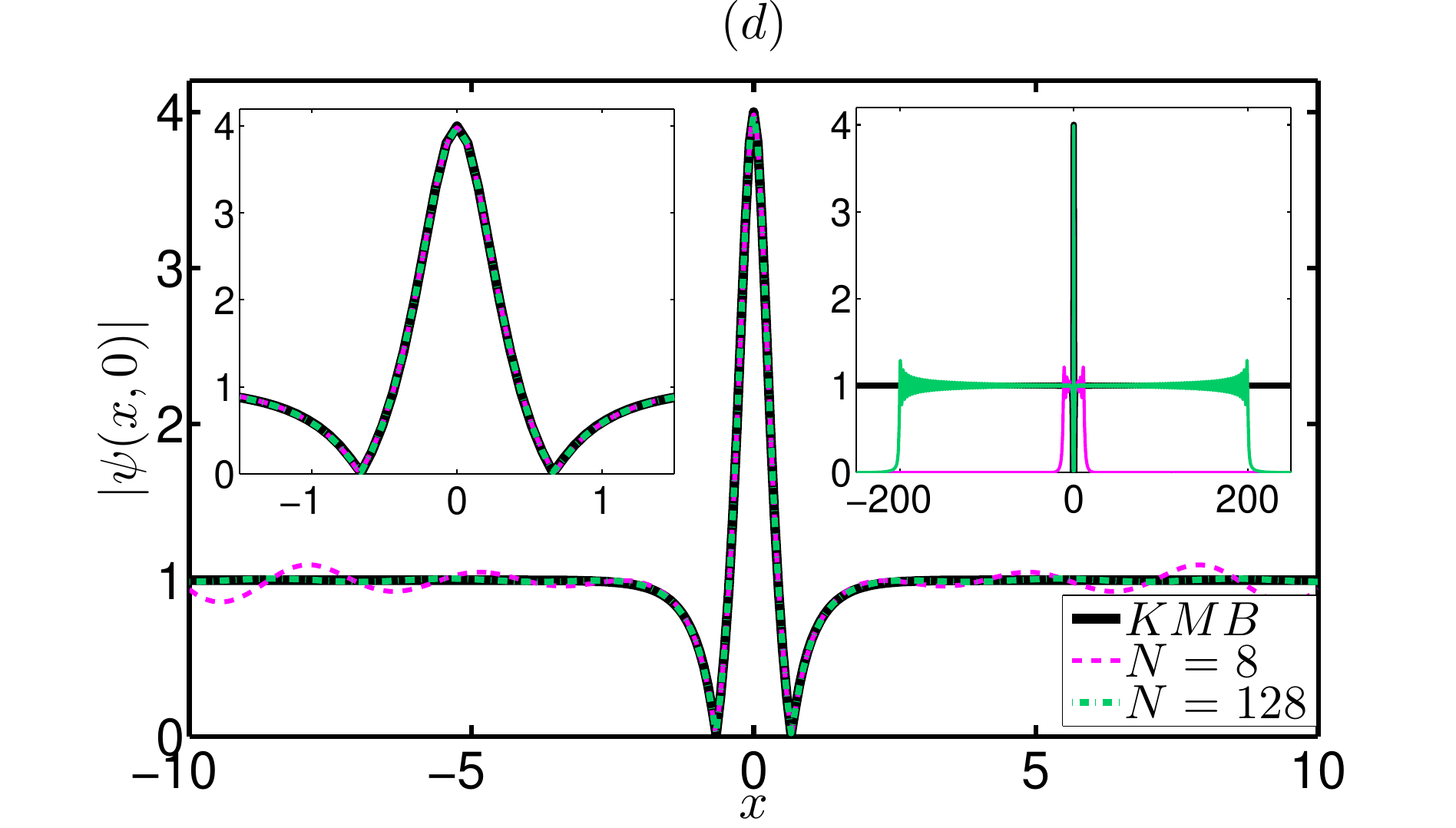}
	\includegraphics[width=0.32\linewidth]{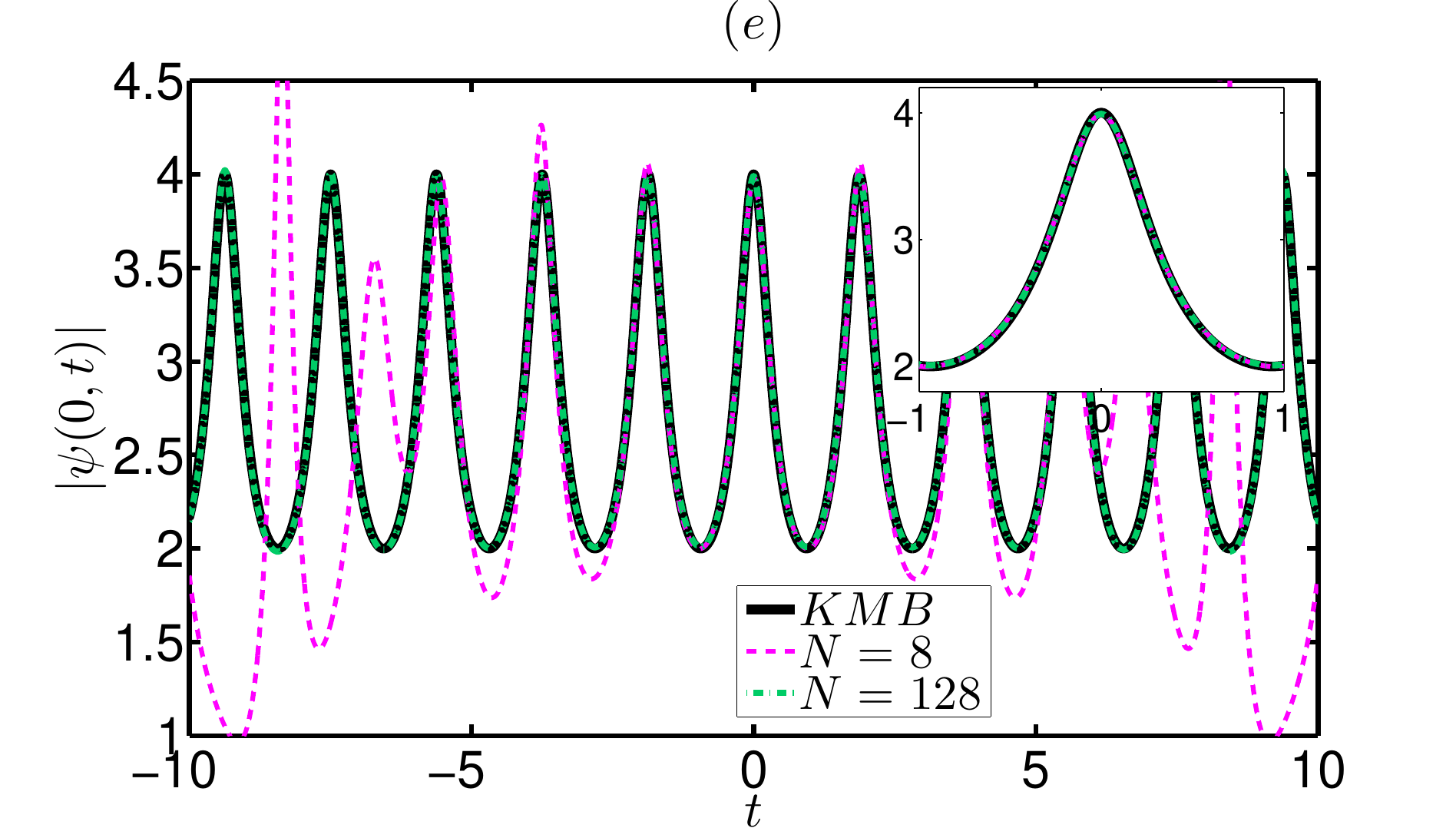}
	\includegraphics[width=0.32\linewidth]{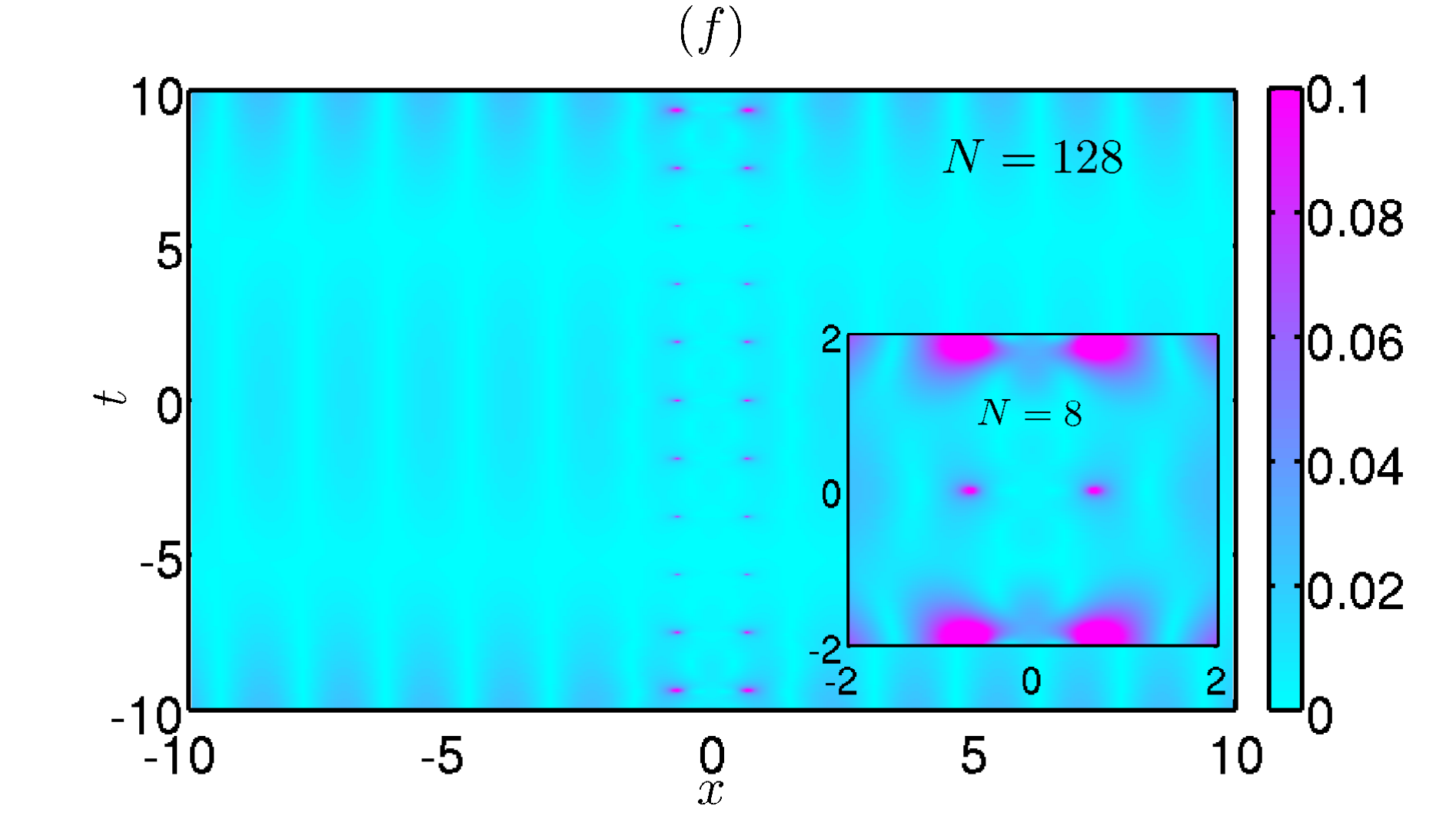}\\
	\includegraphics[width=0.32\linewidth]{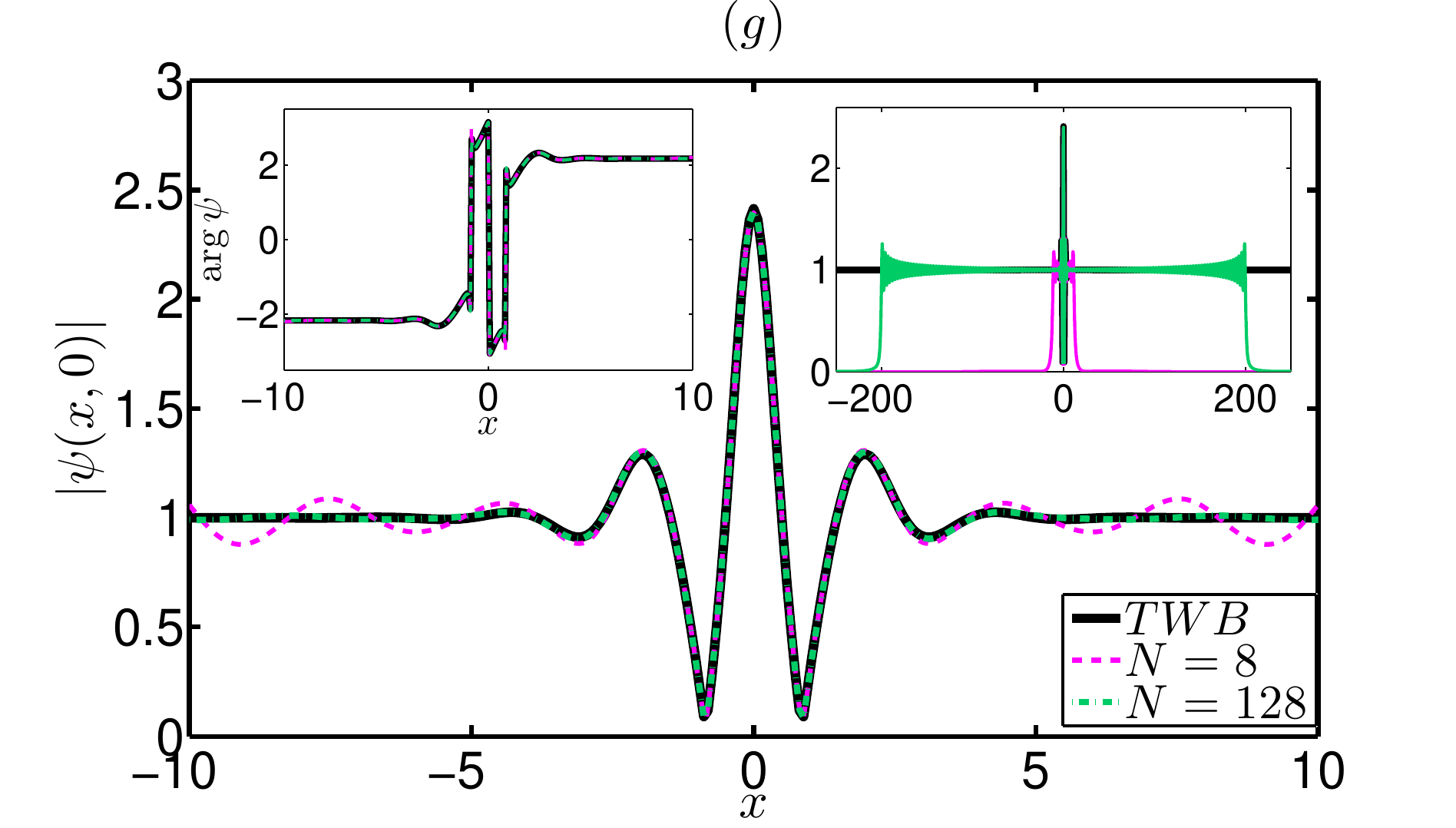}
	\includegraphics[width=0.32\linewidth]{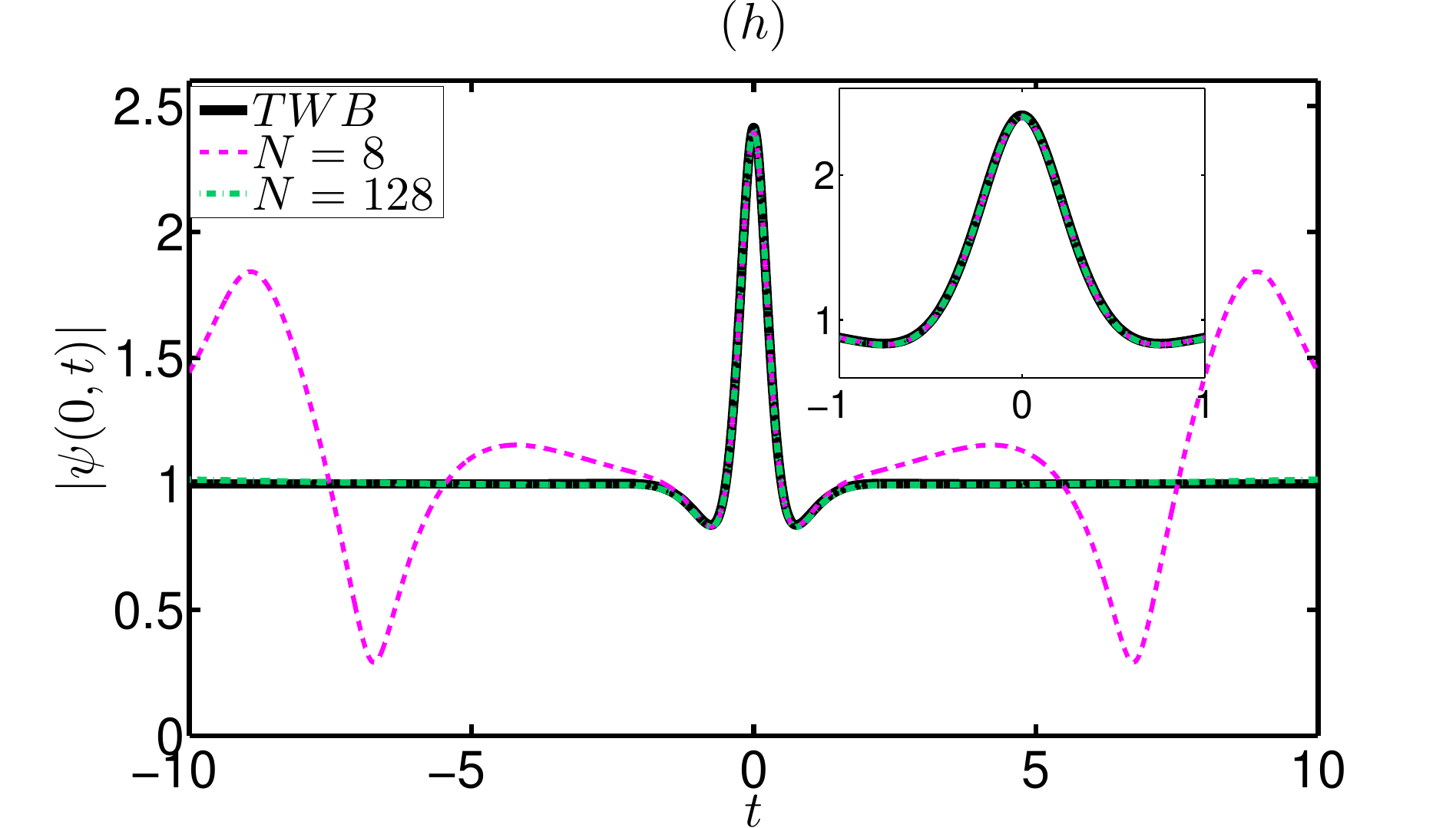}
	\includegraphics[width=0.32\linewidth]{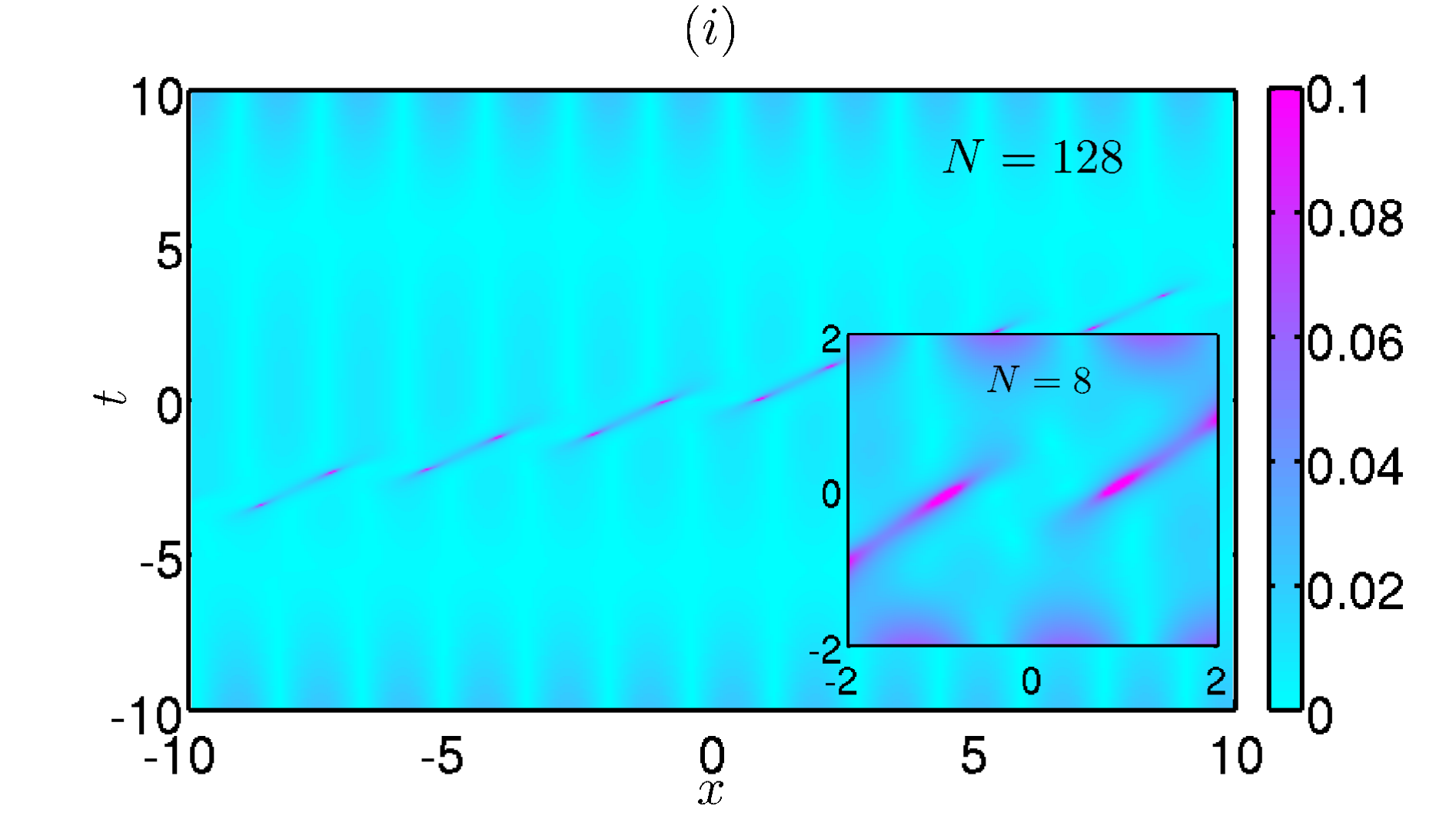}

	\caption{\small {\it (Color on-line)} 
		Same as in Fig.~\ref{fig:fig5}(a,b,c) for the Akhmediev $\lambda_{\mathrm{b}}=i/\sqrt{2}$ (a,b,c), Kuznetsov-Ma $\lambda_{\mathrm{b}}=1.5\,i$ (d,e,f) and Tajiri-Watanabe $\lambda_{\mathrm{b}}=-1 + i/\sqrt{2}$ (g,h,i) breathers.
	}
	\label{fig:figA1}
\end{figure*}

\begin{figure*}[t]\centering
	\includegraphics[width=0.32\linewidth]{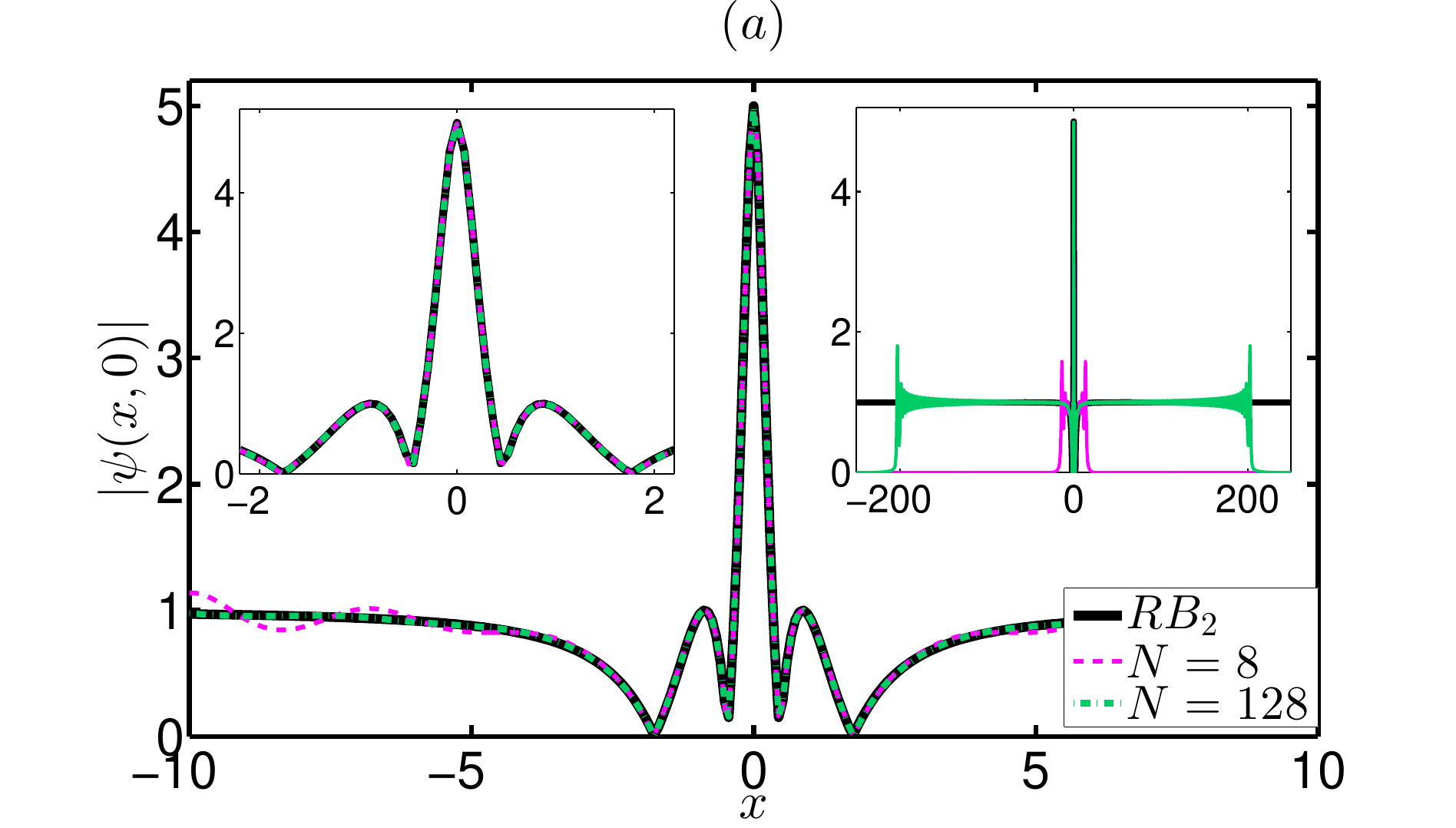}
	\includegraphics[width=0.32\linewidth]{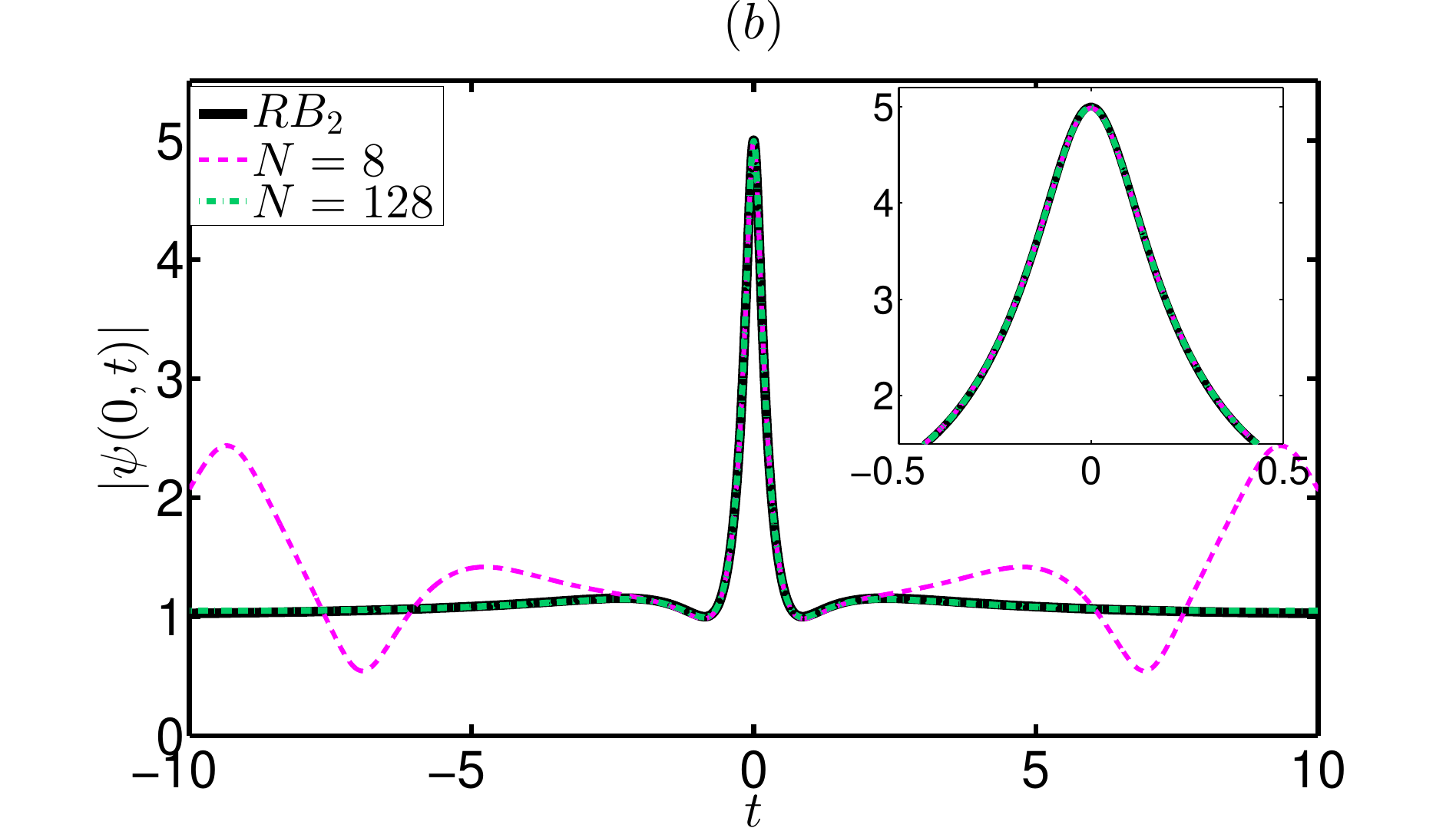}
	\includegraphics[width=0.32\linewidth]{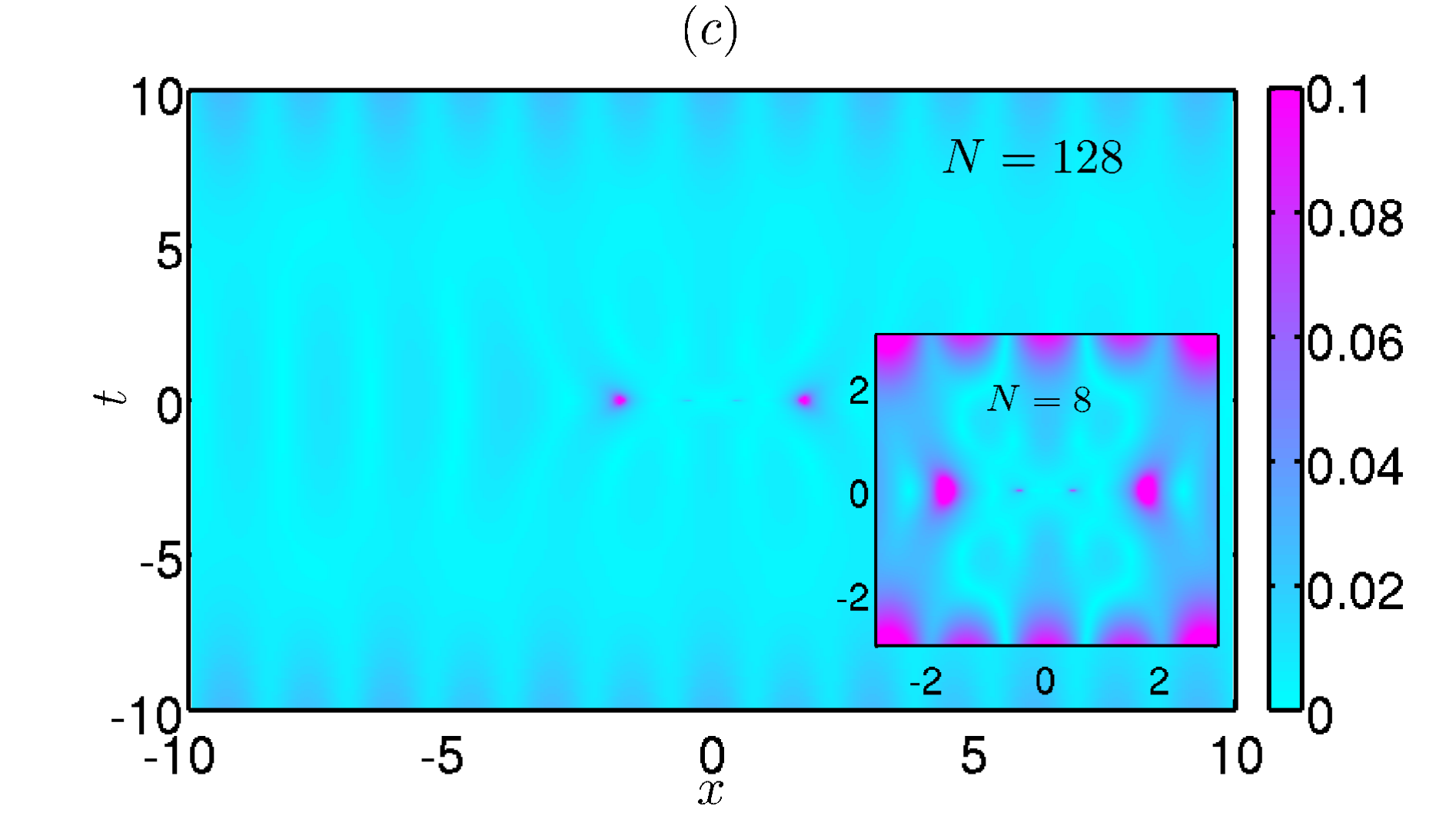}\\
	\includegraphics[width=0.32\linewidth]{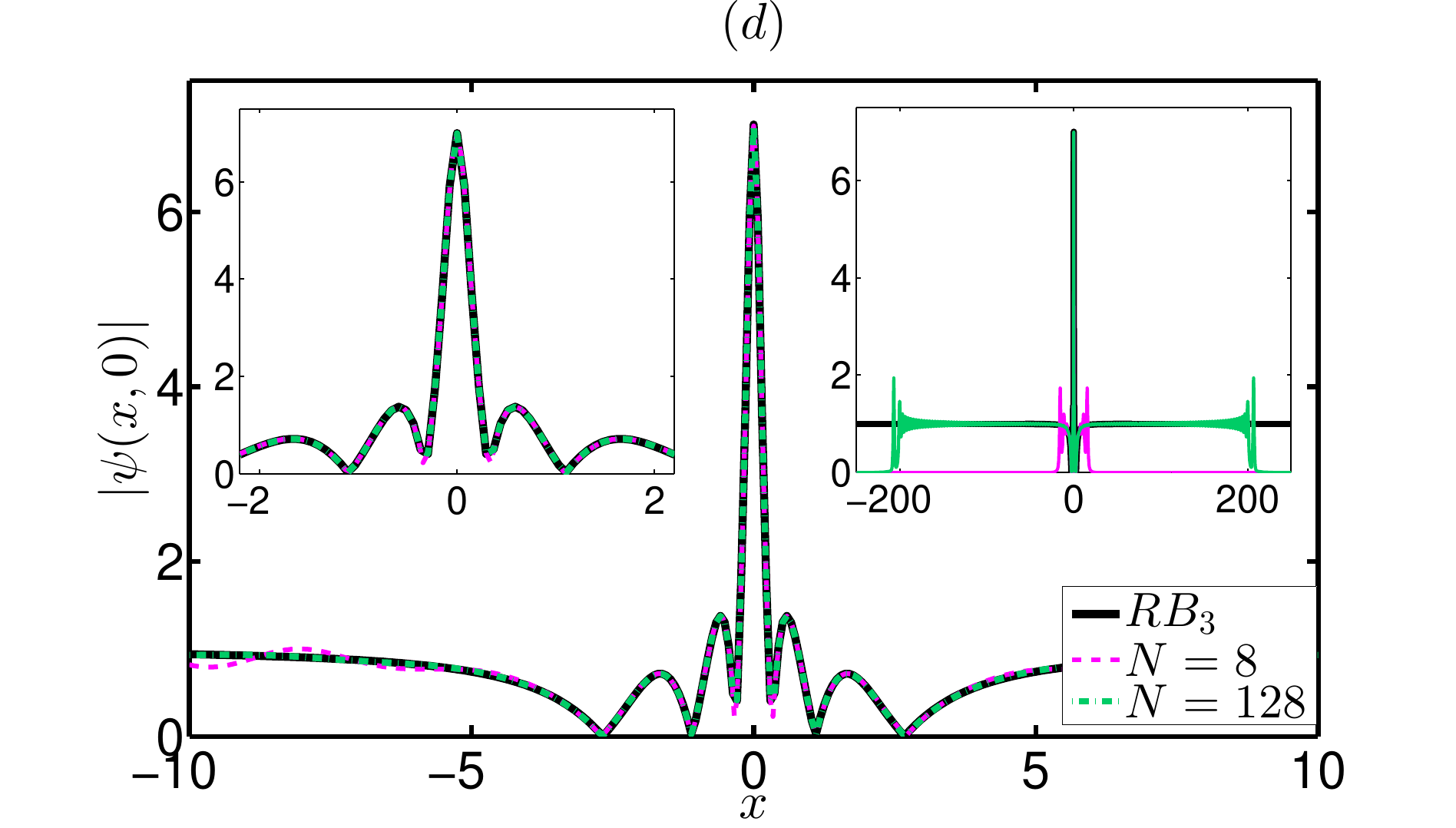}
	\includegraphics[width=0.32\linewidth]{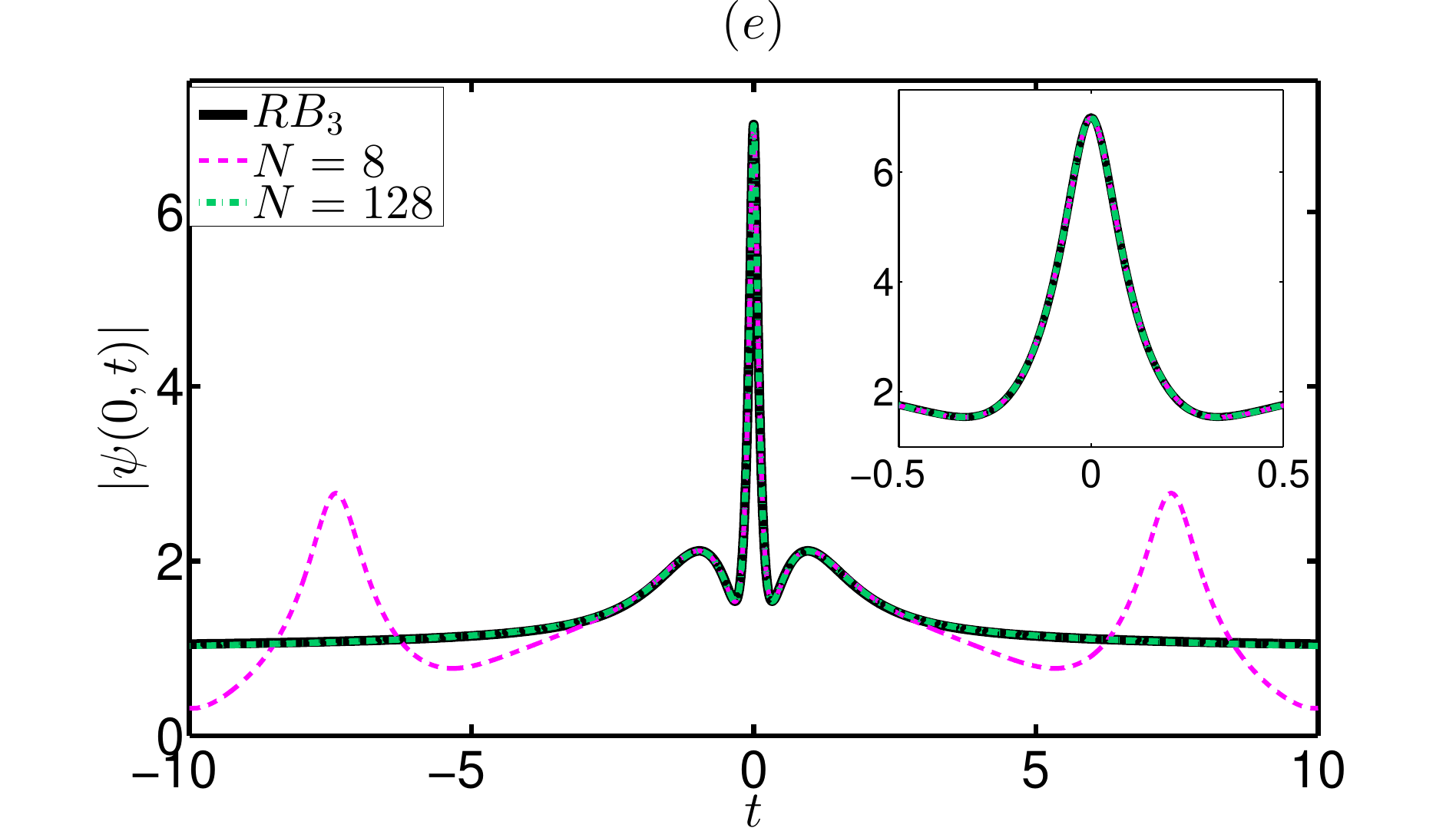}
	\includegraphics[width=0.32\linewidth]{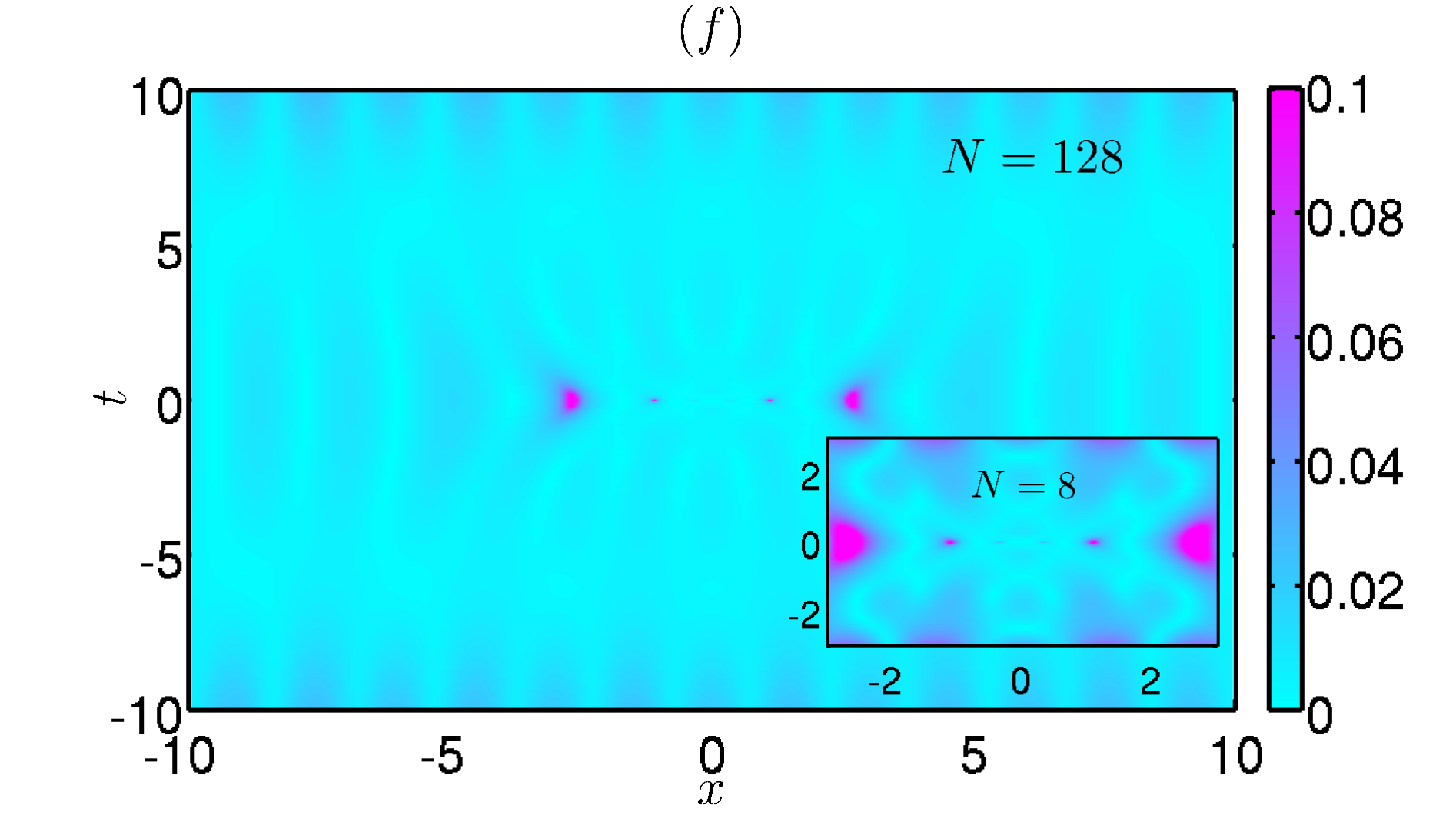}\\
	\includegraphics[width=0.32\linewidth]{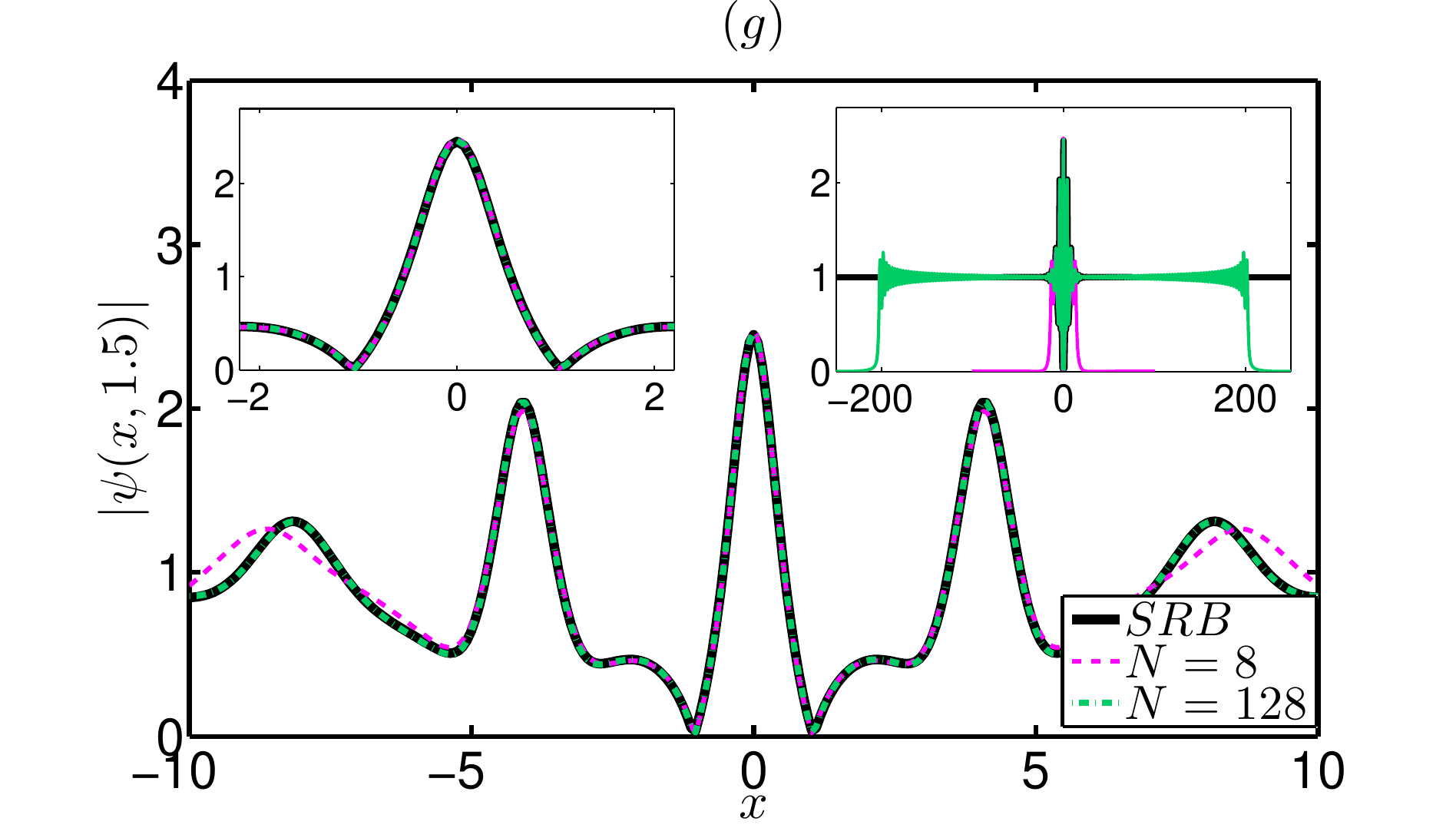}
	\includegraphics[width=0.32\linewidth]{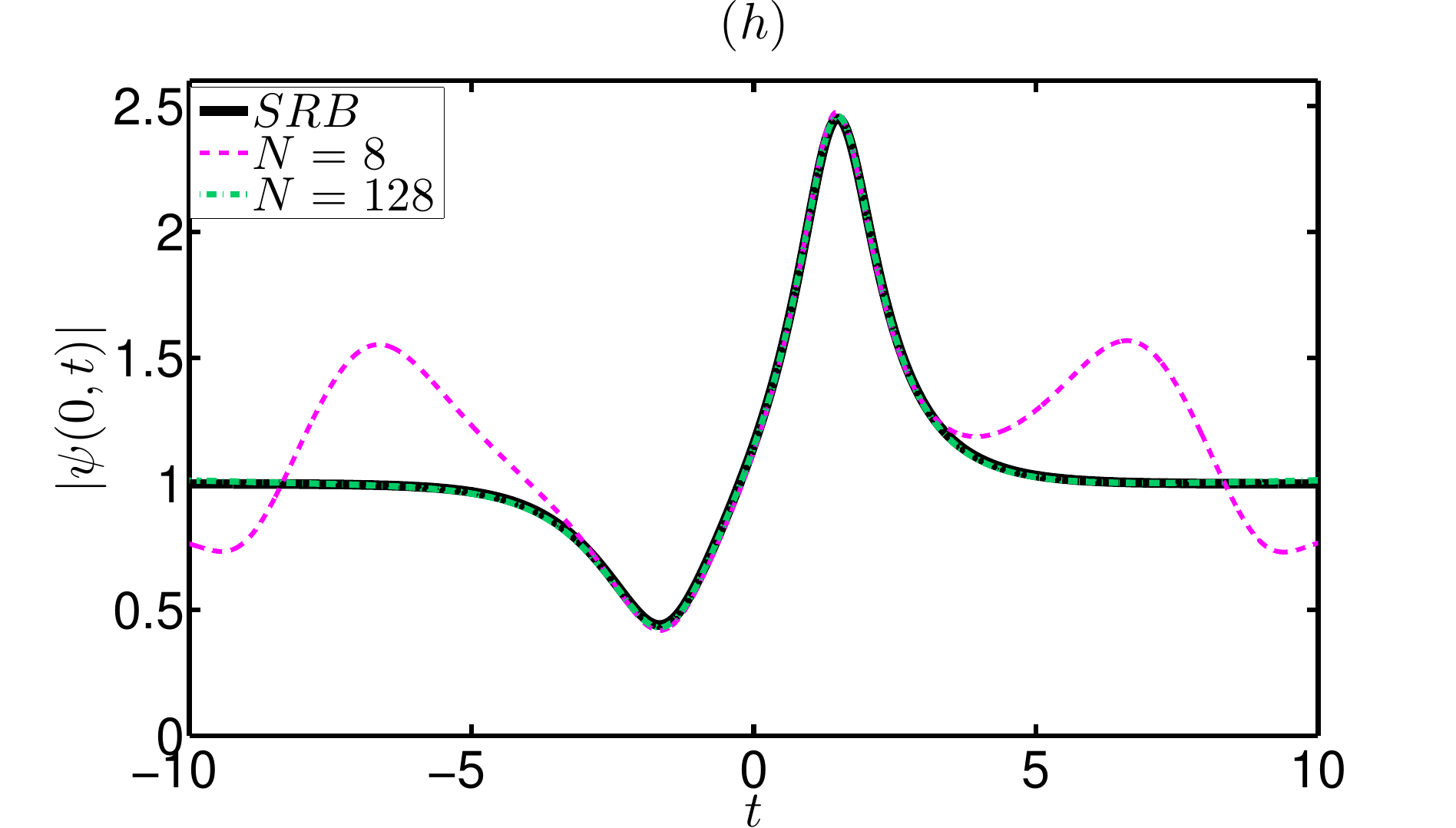}
	\includegraphics[width=0.32\linewidth]{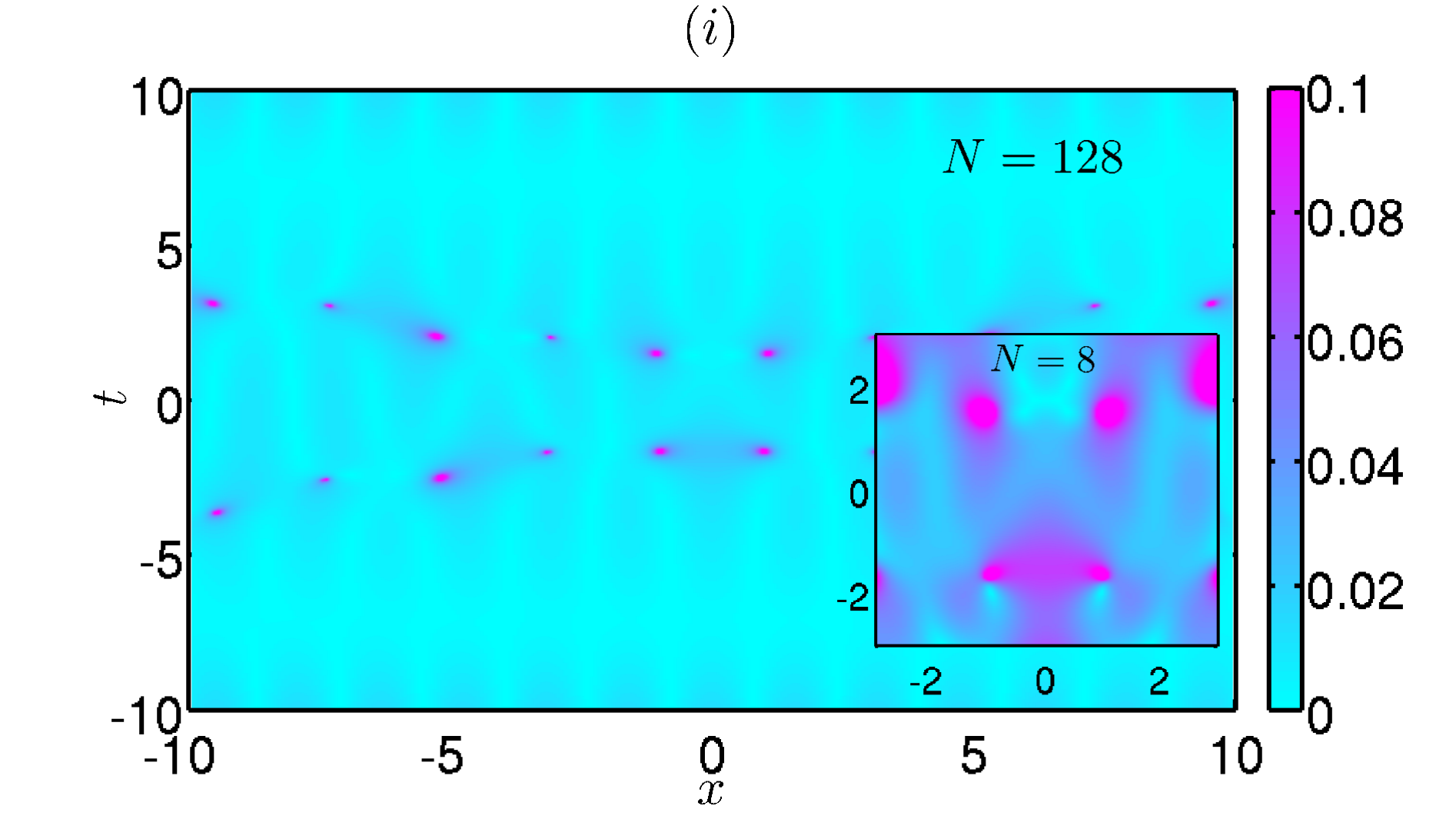}

	\caption{\small {\it (Color on-line)} 
		Same as in Fig.~\ref{fig:fig5}(a,b,c) for the 
		rational breather of the 2nd order $\lambda_{\mathrm{b}1,\mathrm{b}2} = i$ (a,b,c), 
		rational breather of the 3rd order $\lambda_{\mathrm{b}1,\mathrm{b}2,\mathrm{b}3} = i$ (d,e,f), and 
		a pair of super-regular breathers $\lambda_{\mathrm{b}1,\mathrm{b}2} = \pm 0.15 + i/\sqrt{2}$ (g,h,i). 
		The maximum elevation for the super-regular breathers is achieved at $t_{0}\approx 1.5$, so that this moment of time is shown in panel (g) instead of $t=0$.
	}
	\label{fig:figA2}
\end{figure*}

Figure~\ref{fig:figA1}(a,d,g) shows spatial profiles of the breathers at the time $t=0$ of their maximum elevation versus the corresponding $(N+1)$-soliton models for $N=8$ and $N=128$. 
One can see that profiles between the two amplitude minima closest to the coordinate origin $x=0$ are accurately reproduced already by the $9$-soliton solutions ($9$-SS), while the $129$-SS are practically indistinguishable from the breathers over a wide space. 
Comparing the time evolution of the breathers and their solitonic models at $x=0$ in Fig.~\ref{fig:figA1}(b,e,h), we observe that the $9$-SS deviate noticeably at $|t|\gtrsim 2$, while the $129$-SS remain close to the breathers for a much longer time.

The relative deviation~(\ref{deviation-local}) between the breathers and their solitonic models is shown in Fig.~\ref{fig:figA1}(c,f,i): for the $129$-soliton models in the region $(x,t)\in [-10,10]^{2}$ -- in the main figures, and for the $9$-SS in the region $(x,t)\in [-2,2]^{2}$ -- in the insets. 
For all cases, the deviation~(\ref{deviation-local}) remains within $2$\% for most of the areas demonstrated in the figures, and the maximum relative deviation is observed near the local minima $\min|\psi^{\mathrm{B}}(x,t)|$. 
The integral deviation~(\ref{deviation-integral}) is provided in Table~\ref{tab:tab1} in Section~\ref{Sec:Results}; it is about $2$\% for the $9$-soliton models in the region $[-2,2]^{2}$ and about $0.6$\% for the $129$-SS in the region $[-8,8]^{2}$. 

As we have noted in Section~\ref{Sec:Results}, for the case of the Akhmediev breather, the dressing by a single soliton of the solitonic model of the plane wave changes the wavefield of the latter significantly everywhere in space. 
In particular, at $t=0$, the Akhmediev breather $\lambda_{\mathrm{b}}=i/\sqrt{2}$ reaches its maximum $\max|\psi^{\mathrm{A}}|=\sqrt{2}+1$ on the periodic set of points $x = \pi\sqrt{2}\,m$, $m\in\mathbb{Z}$, and this behavior is accurately reproduced by our solitonic model over its entire characteristic width $|x|\lesssim L/2$; see the right inset in Fig.~\ref{fig:figA1}(a). 
The similar situation takes place for the Tajiri-Watanabe breather, for which the dressing procedure changes the phase of the plane wave background to the left and to the right from the breather; see Eq.~(\ref{TW-phase-jump}) and the left inset in Fig.~\ref{fig:figA1}(g).


\section{Higher-order rational and super-regular breathers}
\label{Sec:App:B}

In this Appendix, we generalize our results for specific multi-breather solutions, namely, for the higher-order rational~\cite{akhmediev2009rogue} and super-regular~\cite{zakharov2013nonlinear,gelash2014superregular,kibler2015superregular} breathers. 

\setlength{\tabcolsep}{6pt}
\begin{table}[t]
	\caption{Integral deviations~(\ref{deviation-integral}) between breathers and their solitonic models, for the rational breather of the 2nd order ($\mathrm{RB}_{2}$) $\lambda_{\mathrm{b}1,\mathrm{b}2}=i$, rational breather of the 3rd order ($\mathrm{RB}_{3}$) $\lambda_{\mathrm{b}1,\mathrm{b}2,\mathrm{b}3}=i$, and a pair of super-regular breathers ($\mathrm{SRB}$) $\lambda_{\mathrm{b}1,\mathrm{b}2} = \pm 0.15 + i/\sqrt{2}$.
	}
	
	\begin{center}
		\begin{tabular}{| c | c | c | c |}
			\hline
                \multirow{2}{*}{Breather}	& $N=8$,    & \multicolumn{2}{c}{$N=128$}\vline\\ \cline{3-4}
                                	& $[-2,2]^{2}$ 			& $[-2,2]^{2}$         & $[-8,8]^{2}$           \\ \hline
			$\mathrm{RB}_{2}$ 	& $1.1\times 10^{-2}$ 	& $4.6\times 10^{-3}$  & $6.6\times 10^{-3}$    \\ \hline  
			$\mathrm{RB}_{3}$ 	& $9.4\times 10^{-3}$ 	& $4.0\times 10^{-3}$  & $5.9\times 10^{-3}$    \\ \hline
			$\mathrm{SRB}$		& $3.2\times 10^{-2}$ 	& $7.7\times 10^{-3}$  & $5.7\times 10^{-3}$    \\ \hline
		\end{tabular}
	\end{center}
	\label{tab:tab2}
\end{table}

Higher-order rational breathers are constructed when the (bare) plane wave $\psi_{0}^{\mathrm{B}} = e^{it + i\Theta}$ is repeatedly dressed with the same soliton $\lambda_{\mathrm{b}}=i$ using the same norming constant $C_{\mathrm{b}}$. 
In particular, the second-order rational breather is obtained by double dressing, the third-order breather by triple dressing, and so on. 
Numerically, the dressing scheme discussed in Section~\ref{Sec:Methods:B} cannot be applied in this way due to the division by zero in Eq.~(\ref{qn}) during the repeated dressings, see Eqs.~(\ref{dressing-Psi})-(\ref{dressing-matrix}), while analytically this difficulty can be overcome by applying the L'Hospital's rule. 
For this reason, we build solitonic models of the higher-order rational breathers by using slightly different solitons $\lambda_{\mathrm{b}j}=i(1 + j\,\Delta\eta)$ with the same norming constants $C_{\mathrm{b}j}$, where $\Delta\eta = 10^{-6}$, $j=0,...,M-1$, and $M$ is the order of the breather. 
We use parameters $\Theta = M\pi$ and $C_{\mathrm{b}j} = (-1)^{M-1}$, for which the breathers take their canonical form and represent localized rational perturbations of the plane wave $e^{it}$ leading to the maximum amplitude $\max|\psi^{\mathrm{R}_{M}}(x,t)| = 2 M + 1$ at the point $(0,0)$. 
We do not provide exact analytic relations for these breathers as they are too cumbersome, and instead refer the reader to~\cite{akhmediev2009rogue} where they were first found.

Super-regular breathers are two-breather solutions constructed by dressing with solitons $\lambda_{\mathrm{b}1,\mathrm{b}2}=\pm\xi_{\mathrm{b}} + i\eta_{\mathrm{b}}$ with $|\xi_{\mathrm{b}}|\ll 1$ and $\eta_{\mathrm{b}}<1$. 
For specific combinations of the norming constants $C_{\mathrm{b}1,\mathrm{b}2}$, these solutions represent localized perturbations of the plane wave with characteristic amplitude $\propto |\xi_{\mathrm{b}}|$ and width $\propto |\xi_{\mathrm{b}}|^{-1}$, which lead to the same phases of the plane wave background in the limits $x\to\pm\infty$; see Eq.~(\ref{N-phase-jump}). 
We build their solitonic models using the parameters $\Theta=0$ and $C_{\mathrm{b}1,\mathrm{b}2}=-i$, for which the breathers represent small localized perturbations of the plane wave $e^{it}$, and compare our models with the exact solutions first found in~\cite{zakharov2013nonlinear}. Note that a more general choice of eigenvalues is possible~\cite{gelash2014superregular,gelash2018formation} with non-symmetric spatial profiles of the wavefields. 
For definiteness, we use $\lambda_{\mathrm{b}1,\mathrm{b}2}=\pm 0.15 + i/\sqrt{2}$; we have tried other values and came to the same results.

Figure~\ref{fig:figA2} shows comparison of our solitonic models for $N=8$ and $128$ with the exact breather solutions for the second-order rational (a,b,c), third-order rational (d,e,f), and super-regular (g,h,i) breathers. 
Here $N$ is the number of solitons used for modeling the plane wave background, so that our solitonic models contain $10$-$11$ and $130$-$131$ solitons. 
One can see that the results of this comparison are very similar to those discussed in Section~\ref{Sec:Results} and in Appendix~\ref{Sec:App:A}. 
In particular, the spatial profiles of breathers at the time of their maximum elevation are accurately reproduced already by the $N=8$ models between the two adjacent local minima, while the $N=128$ models are practically indistinguishable from breathers over a wide region of space and time. 
The integral deviation~(\ref{deviation-integral}) between our solitonic models and breathers given in Table~\ref{tab:tab2} equals $1$-$3$\% for the $N=8$ models in the region $[-2,2]^{2}$ and about $0.6$\% for the $N=128$ models in the region $[-8,8]^{2}$. 



\begin{figure*}[t]\centering
    \includegraphics[width=0.47\linewidth]{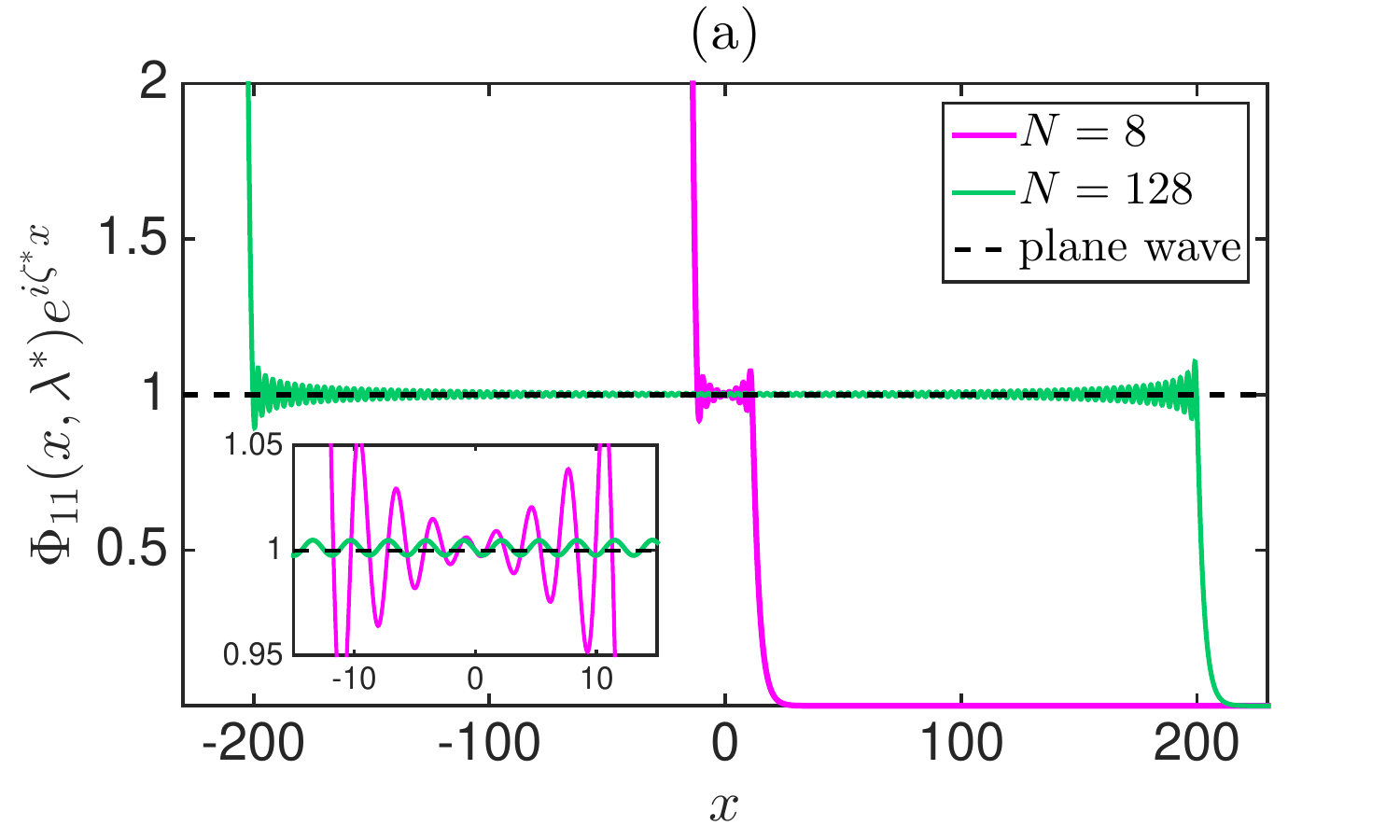}
    \includegraphics[width=0.47\linewidth]{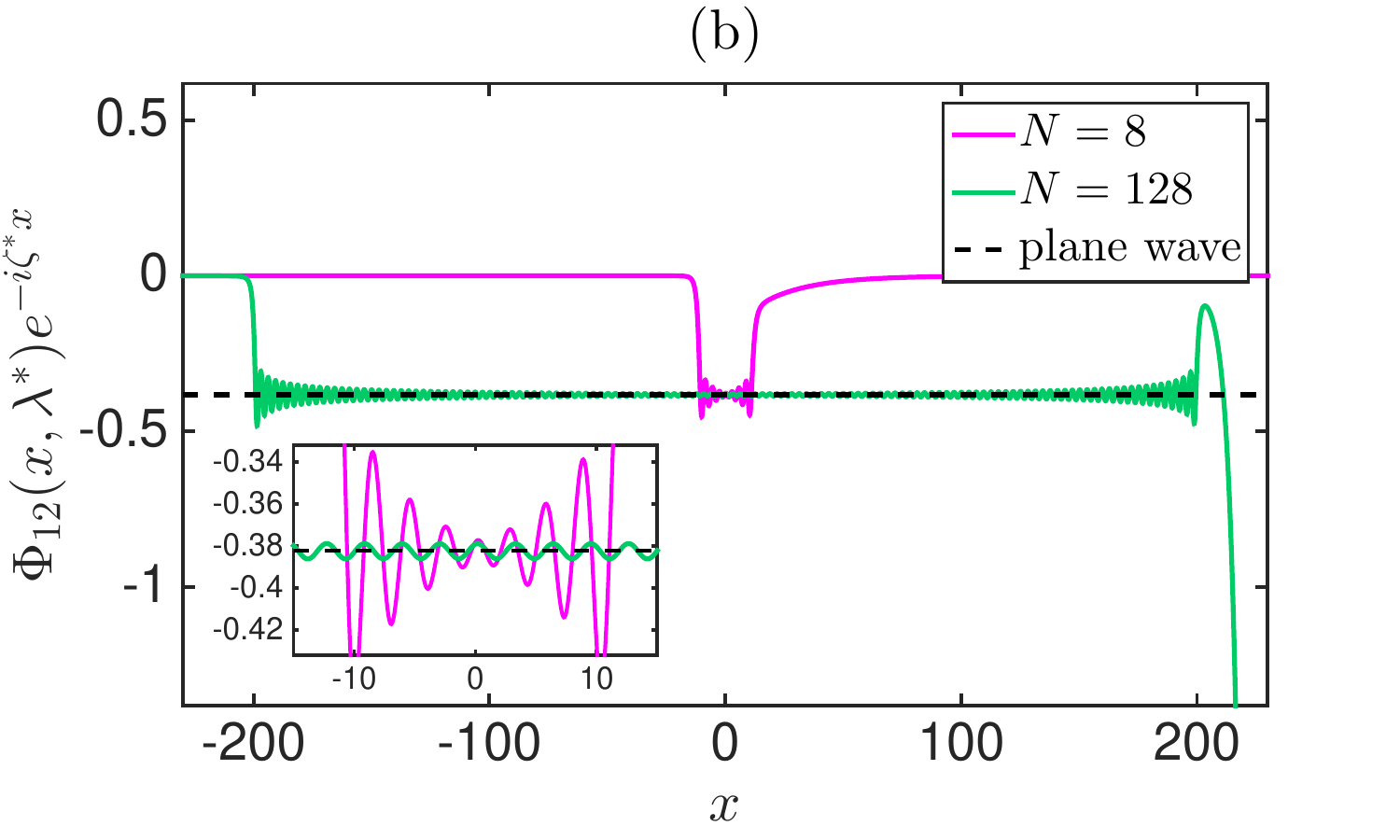}\\
    \includegraphics[width=0.47\linewidth]{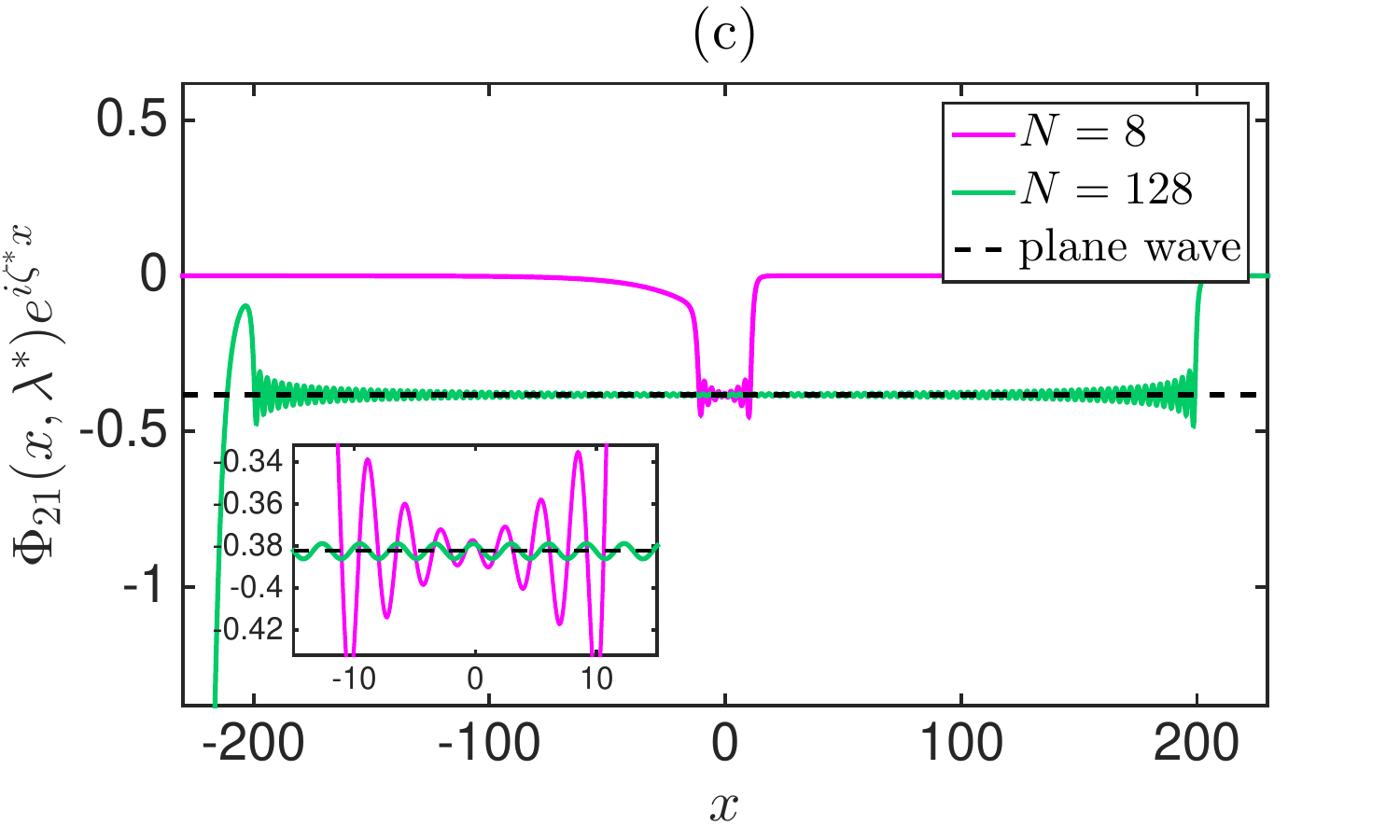}
    \includegraphics[width=0.47\linewidth]{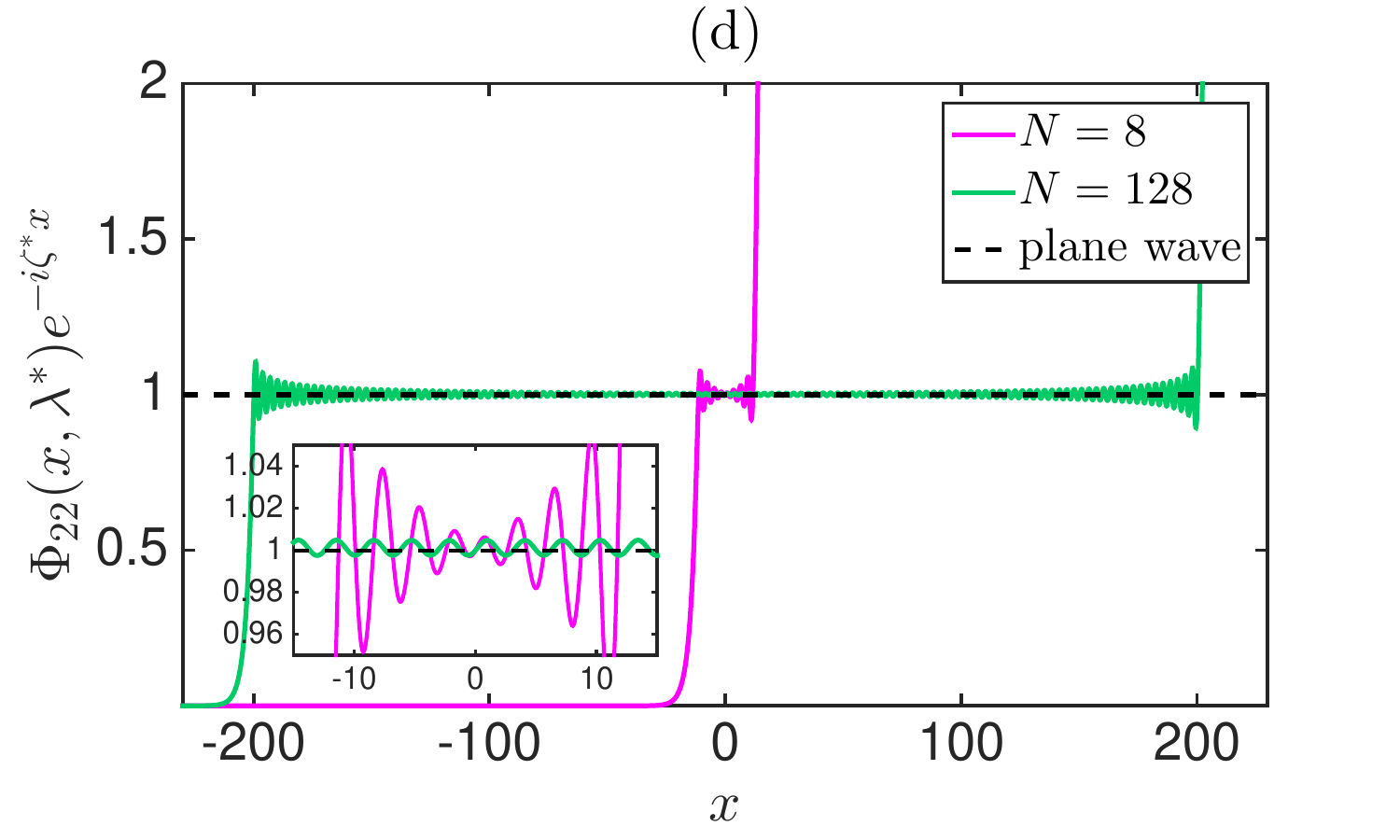}
    
    \caption{\small {\it (Color on-line)} 
    Spatial behavior of the elements of wave functions $\mathbf{\Phi}_{0}^{\mathrm{B}}(x,t,\lambda^{*})$ (dashed black) and $\mathbf{\Phi}_{N}^{\mathrm{S}}(x,t,\lambda^{*})$ (magenta for $N=8$ and green for $N=128$) at $t=0$ and $\lambda=1.5\,i$: (a) $\Phi_{11}$, (b) $\Phi_{12}$, (c) $\Phi_{21}$ and (d) $\Phi_{22}$. 
    As in Section~\ref{Sec:Results}, we use $\Theta=\pi$ for the plane wave and its solitonic model. 
    Both wave functions are normalized so that their first elements equal unity at $x=0$. 
    For visual comparison, we multiply $\Phi_{ij}$, $i,j=1,2$, either by $e^{i\zeta^* x}$, or by $e^{-i\zeta^* x}$, to compensate the exponential behavior. 
    We demonstrate only the real part of wave functions, since the imaginary part is identically zero at $t=0$. 
    }
    \label{fig:figA3}
\end{figure*}

\section{Wave function for the solitonic model of the plane wave}
\label{Sec:App:C}

As follows from Eqs.~(\ref{psi_n})-(\ref{dressing-matrix}), the dressed potential $\psi_{n}$ and wave function $\mathbf{\Phi}_{n}$ depend on $\psi_{n-1}$ and $\mathbf{\Phi}_{n-1}$ taken at the same point $(x,t)$, so that the dressing procedure is local. 
Also, if two bare potentials $\psi_{0}$ and $\tilde{\psi}_{0}$ are similar, and the corresponding wave functions $\mathbf{\Phi}_{0}$ and $\tilde{\mathbf{\Phi}}_{0}$ are similar as well, then their identical dressing will lead to the similar potentials $\psi_{1}$ and $\tilde{\psi}_{1}$, meaning the continuity of the dressing procedure. 
This property may underlie our results on the similarity between breather solutions and our solitonic models -- if only the wave functions $\mathbf{\Phi}_{0}^{\mathrm{B}}$ and $\mathbf{\Phi}_{N}^{\mathrm{S}}$ corresponding to the plane wave and its $N$-soliton model are similar. 
Here we verify this similarity numerically by comparing wave function~(\ref{Psi0cond}) with that for the solitonic model of the plane wave. 

Note that the wave function depends on three variables $x$, $t$ and $\lambda$. 
We are unable to perform a systematic comparison in this three-dimensional space, and instead fix $t=0$ and $\lambda=1.5\,i$, and then compare the spatial profiles over the $x$-coordinate. 
We have tried a number of other combinations of times $t$ (not large) and eigenvalues $\lambda$ (both $\mathrm{Im}\,\lambda\ge 1$ and $\mathrm{Im}\,\lambda<1$), and came to the similar results. 
Also note that, in the dressing procedure, the wave function of the previous step is evaluated at the complex-conjugate point $\lambda^{*}$, see Eq.~(\ref{qn}). 
For this reason, in the following, we perform comparison between wave functions $\mathbf{\Phi}_{0}^{\mathrm{B}}(x,t,\lambda^{*})$ and $\mathbf{\Phi}_{N}^{\mathrm{S}}(x,t,\lambda^{*})$. 

If $\lambda$ does not belong to the discrete spectrum of the solitonic model of the plane wave~(\ref{box_eigenvalues}), then both $\mathbf{\Phi}_{0}^{\mathrm{B}}(x,t,\lambda^{*})$ and $\mathbf{\Phi}_{N}^{\mathrm{S}}(x,t,\lambda^{*})$ are unbounded and exhibit exponential behavior in the $x$-space, see e.g. Eq.~(\ref{Psi0cond}). 
For visual comparison, we multiply the elements of these matrices either by $e^{i\zeta^{*} x}$, or by $e^{-i\zeta^{*} x}$, compensating the leading exponent. 
Additionally, the ZS system is linear, so that different solutions may have different normalizations. 
For this reason, before comparison, we normalize wave functions by multiplying all four matrix elements by the same constant so that the first matrix element becomes unity at $x=0$, $\Phi_{11}(0,t,\lambda^{*})=1$. 
As follows from Eqs.~(\ref{psi_n})-(\ref{qn}) and~(\ref{dressing-matrix}), such a change of normalization does not affect the dressing procedure. 

Figure~\ref{fig:figA3} shows elements of wave function~(\ref{Psi0cond}) (dashed black lines) in comparison with the corresponding elements of $\mathbf{\Phi}^{\mathrm{S}}_N$ for $N=8$ (magenta) and $N=128$ (green). 
One can see a very good agreement over the entire characteristic width $|x|\lesssim L/2$, $L=\pi(N-1/4)$, of the solitonic model of the plane wave. 
Very similarly to this model (compare with Fig.~\ref{fig:fig3}), $\mathbf{\Phi}^{\mathrm{S}}_N$ oscillates around the exact solution~(\ref{Psi0cond}), and these oscillations decrease with the growing number of solitons $N$. 

This comparison confirms that our solitonic model of the Kuznetsov-Ma breather $\lambda_{\mathrm{b}}=1.5\,i$ is similar to the exact breather due to (i) the locality and continuity of the dressing procedure, and (ii) the fact that the wave function $\mathbf{\Phi}^{\mathrm{S}}_N$ corresponding to the solitonic model of the plane wave is similar to the exact solution~(\ref{Psi0cond}) at $\lambda^{*}=-1.5\,i$. 
We have come to the same result for a number of other eigenvalues, including those corresponding to the Akhmediev breather.


%

\end{document}